\documentclass[11pt,a4paper]{article}
\usepackage[latin1]{inputenc}
\usepackage{amsmath}
\usepackage{amsfonts}
\usepackage{amssymb}
\usepackage{amsthm}
\usepackage{graphicx}
\usepackage[round,authoryear]{natbib}
\usepackage{setspace}
\usepackage{footmisc}
\usepackage[english]{babel}
\usepackage{authblk}
\usepackage[margin=3cm]{geometry}
\usepackage{mathtools}
\usepackage{enumitem}
\usepackage{thmtools}
\usepackage{thm-restate}

\usepackage{xcolor}

\newif\ifanon
\anonfalse

\ifanon \else
\usepackage{hyperref}
\fi
\usepackage{cleveref}

\newcommand\sdArgs[2]{#1 \succ_{sd} #2}
\newcommand\cdfDiff[2]{\Delta_{#1#2}}
\newcommand*\sd{\succ_{sd}}

\newcommand*\ccdf[1]{\bar{#1}}
\newcommand*\bupdf{\beta}
\newcommand*\bucdf{B}
\newcommand*\arbitrarypayoff{v}

\DeclareMathOperator{\maxRatioParam}{rate}
\DeclareMathOperator\supp{supp}
\DeclareMathOperator\IQR{IQR}

\ifanon
\title{Exceeding Expectations: Stochastic Dominance as the Criterion of Rational Choice}
\author{}
\date{}

\else
\title{Exceeding Expectations: Stochastic Dominance as a General Decision Theory}
\author{Christian J.\@ Tarsney\thanks{Global Priorities Institute, Faculty of Philosophy, University of Oxford}}
\date{\normalsize Version 8, August 2020}
\fi

\begin{document}
\maketitle

%
%

\begin{abstract}
The principle that rational agents should maximize expected utility or choiceworthiness is intuitively plausible in many ordinary cases of decision-making under uncertainty. But it is less plausible in cases of extreme, low-probability risk (like Pascal's Mugging), and intolerably paradoxical in cases like the St.\@ Petersburg and Pasadena games. In this paper I show that, under certain conditions, stochastic dominance reasoning can capture most of the plausible implications of expectational reasoning while avoiding most of its pitfalls. Specifically, given sufficient background uncertainty about the choiceworthiness of one's options, many expectation-maximizing gambles that do not stochastically dominate their alternatives `in a vacuum' become stochastically dominant in virtue of that background uncertainty. But, even under these conditions, stochastic dominance will not require agents to accept options whose expectational superiority depends on sufficiently small probabilities of extreme payoffs. 
The sort of background uncertainty on which these results depend looks unavoidable for any agent 
who measures the choiceworthiness of her options in part by the total amount of value in the resulting world. 
At least for such agents, then, stochastic dominance offers a plausible general principle of choice under uncertainty that can explain more of the apparent rational constraints on such choices than has previously been recognized. \ifanon [Word count: 19,265 (15,282 main text + 3983 notes)] \else \fi 
\end{abstract}


\section{Introduction}
\label{section-intro}



Given our epistemic limitations, every choice you or I will ever make involves some degree of risk. Whatever we do, it might turn out that we would have done better to do something else. If our choices are to be more than mere leaps in the dark, therefore, we need principles that tell us how to evaluate and compare risky options.



The standard view in normative decision theory holds we should rank options by their \textit{expectations}. That is, an agent should assign cardinal degrees of utility or choiceworthiness to each of her available options in each possible state of nature, assign probabilities to the states, and prefer one option to another just in case the probability-weighted sum (i.e., expectation) of its possible degrees of utility or choiceworthiness is greater. 
Call this view \textit{expectationalism}.


Expectational reasoning provides seemingly indispensable practical guidance in many ordinary cases of decision-making under uncertainty.\footnote{Throughout the paper, I assume that agents can assign precise probabilities to all decision-relevant possibilities. 
Since there is little possibility of confusion, therefore, I use `risk' and `uncertainty' interchangeably, setting aside the familiar distinction due to \cite{knight1921risk}. I default to `uncertainty' (and, in particular, speak of `background uncertainty' rather than `background risk') partly to avoid the misleading connotation of `risk' as something exclusively negative.} But it encounters serious difficulties in many cases involving extremely large finite or infinite payoffs, where it yields conclusions that are either implausible, unhelpful, or both. For instance, expectationalism implies that: (i) Any positive probability of an infinite positive or negative payoff, no matter how minuscule, takes precedence over all finitary considerations (\citeauthor{pascal1852pensees}, 1669). 
(ii) When two options carry positive probabilities of infinite payoffs of the same sign (i.e., both positive or both negative), and zero probability of infinite payoffs of the opposite sign, the two options are 
equivalent, even if one offers a much greater probability of that infinite payoff than the other \citep{hajek2003waging}. (iii) When an option carries any positive probabilities of both infinite positive and infinite negative payoffs, it is simply incomparable with any other option 
\citep{bostrom2011infinite}. (iv) Certain probability distributions over finite payoffs yield expectations that are infinite (as in the St.\@ Petersburg game (\citeauthor{bernoulli1954exposition}, 1738)) or undefined (as in the Pasadena game \citep{nover2004vexing}), so that options with these prospects are better than or incomparable with any guaranteed finite payoff.\footnote{As is common in discussions of the St.\@ Petersburg game, I assume here that we can extend the strict notion of an expectation to allow that, when the probability-weighted sum of possible payoffs diverges unconditionally to $+/- \infty$, the resulting expectation is infinite rather than undefined.} 
(v) Agents can be rationally required to prefer minuscule probabilities of astronomically large finite payoffs over certainty of a more modest payoff, in cases where that preference seems at best rationally optional (as in `Pascal's Mugging' \citep{bostrom2009pascal}).



The last of these problem cases, though theoretically the most straightforward, has particular practical significance. 
Real-world agents 
who want to do the most good when they choose a career or donate to charity often face choices between options that are fairly likely to do a moderately large amount of good (e.g., supporting public health initiatives in the developing world or promoting farm animal welfare) and options that carry much smaller probabilities 
of doing much larger amounts of good (e.g., reducing existential risks to human civilization \citep{bostrom2013existential,ord2020precipice} or trying to bring about very long-term `trajectory changes' \citep{beckstead2013overwhelming}). Often, na\"{i}ve application of expectational reasoning suggests that we are rationally required to choose the latter sort of project, even if the probability of having any positive impact whatsoever 
is vanishingly small.  For instance, based on 
an estimate that future Earth-originating civilization might support the equivalent of $10^{52}$ human lives, Nick Bostrom concludes that, `[e]ven if we give this allegedly lower bound...
a mere 1 per cent chance of being correct, we find that the expected value of reducing existential risk by a mere \textit{one billionth of one billionth of one percentage point} is worth a hundred billion times as much as a billion human lives' \cite[p.\@ 19]{bostrom2013existential}. This 
suggests that we should 
pass up opportunities to do enormous amounts of good 
in the present, 
to maximize 
the probability of an astronomically good future, 
even if the probability of having any effect at all
is on the order of,  say, $10^{-30}$---meaning, for all intents and purposes, \textit{no matter how small} the probability.

Even hardened utilitarians who think we should normally do what maximizes expected welfare may 
find this conclusion troubling and counterintuitive. We intuit (or so I take it) not that the expectationally superior long-shot option is \textit{irrational}, but simply that it is \textit{rationally optional}: We are not rationally required to forego a high probability of doing a significant amount of good for a vanishingly small probability of doing astronomical amounts of good. And we would like decision theory to vindicate this judgment.




The aim of this paper is to set out an alternative to expectational decision theory that outperforms it in the various problem cases just described---but in particular, with respect to tiny probabilities of astronomical payoffs. Specifically, I will argue that under plausible epistemic conditions, \textit{stochastic dominance reasoning} can capture most of the ordinary, attractive implications of expectational decision theory---far more 
than has previously been recognized---while avoiding its pitfalls in the 
problem cases described above, and in particular, while permitting us to decline expectationally superior options in extreme, `Pascalian' choice situations.


Stochastic dominance is, on its face, an extremely modest principle of rational choice, simply formalizing the idea that one ought to prefer a given probability of a better payoff to the same probability of a worse payoff, all else being equal. The claim that we are rationally required to reject stochastically dominated options is therefore on strong \textit{a priori} footing (considerably stronger, I will argue, than expectationalism). But precisely because it is so modest, stochastic dominance seems too weak to serve as a final principle of decision-making under uncertainty: It appears to place no constraints on an agent's \textit{risk attitudes}, allowing intuitively irrational extremes of risk-seeking and risk-aversion.


But in fact, 
stochastic dominance has a 
hidden capacity to effectively constrain 
risk attitudes: When an agent is in a state of sufficient `background uncertainty' about the choiceworthiness of her options, expectationally superior options that would not otherwise stochastically dominate their alternatives can become stochastically dominant. 
Background uncertainty generates stochastic dominance much less readily, however, in situations where the balance of expectations is determined by minuscule probabilities of astronomical positive or negative payoffs. 
Stochastic dominance thereby draws a principled line between `ordinary' and `Pascalian' choice situations, and vindicates our intuition that we are often permitted to decline 
gambles like Pascal's Mugging or the St.\@ Petersburg game, even when they are expectationally best. Since it avoids these and other pitfalls of expectational reasoning, if stochastic dominance can also place plausible constraints on our risk attitudes and thereby recover the attractive implications of expectationalism, it may provide a more attractive 
criterion of rational choice under uncertainty.

I begin in \S \ref{section-exepctationalism} by saying more about 
standard expectational decision theory, as motivation and point of departure for my main line of argument. 
\S \ref{section-stochasticDominance} introduces stochastic dominance. 
\S \ref{section-formalSetup} gives a formal framework for describing decisions under background uncertainty. In \S \ref{section-results}, I establish two central results: (i) a sufficient condition for stochastic dominance which implies, among other things, that whenever $O_i$ is expectationally superior to $O_j$, it will come to stochastically dominate $O_j$ given sufficient background uncertainty; 
and (ii) a necessary condition for stochastic dominance which implies, among other things, that it is harder for expectationally superior options to become stochastically dominant under background uncertainty when their expectational superiority depends on small probabilities of extreme payoffs. 
In \S \ref{section-sourcesofbackgrounduncertainty}, I argue that the sort of background uncertainty on which these results depend is rationally appropriate at least for any agent who assigns normative weight to \textit{aggregative consequentialist} considerations, i.e., who measures the choiceworthiness of her options at least in part by the total amount of value in the resulting world. \S \ref{section-relevanceofbackgrounduncertainty} offers an intuitive defense of the initially implausible conclusion that an agent's background uncertainty can make a difference to what she is rationally required to do. \S \ref{section-twoModestConclusions} describes two modest conclusions we might draw from the preceding arguments, short of embracing stochastic dominance as a sufficient criterion of rational choice. In \S \ref{section-stochasticdominancedecisiontheory}, however, I survey several further advantages of stochastic dominance over expectational reasoning and argue that, insofar as stochastic dominance can recover 
the intuitively desirable implications of expectationalism, we have substantial reason to prefer it as a 
criterion of rational choice under uncertainty. \S \ref{section-conclusion} is the conclusion.


\section{Expectationalism}
\label{section-exepctationalism}

\subsection{Preliminaries}

Practical rationality (hereafter, `rationality') involves responding correctly to one's beliefs about practical reasons.\footnote{I don't claim that this is all there is to practical rationality---some rational requirements, like the requirement against forming inconsistent intentions, may have a different source. 
But the decision-theoretic aspect of practical rationality with which this paper is concerned does, I assume, consist in responding correctly to reason-beliefs.} 
Following others in the recent literature (e.g.\@ \cite{wedgwood2013akrasia,wedgwood2017must}, \cite{lazar2017deontological}, \cite{macaskill2020why}), I will speak of the total, all-things-considered strength of an agent's reasons for or against choosing a particular option as the \textit{choiceworthiness} of that option. Reasons and choiceworthiness, in the sense we're concerned with, are objective in the sense of being `fact-relative' rather than `belief-relative' \citep{parfit2011onwhatmatters}---e.g., the fact that my glass is poisoned gives me a reason against drinking from it, and thereby makes the option of drinking less choiceworthy, even if I neither believe nor have any evidence that it is poisoned. I take no stance on whether an option's choiceworthiness depends on the agent's motivational states (desires, preferences, etc), on acts of will (e.g.\@ willing certain ends for herself), or on external normative/evaluative features of the world (e.g.\@ universal moral obligations). In other words, choiceworthiness is objective in the sense of being \textit{belief}-independent, but may or may not be objective in the sense of being \textit{desire}- or \textit{preference}-independent.\footnote{I use the term `choiceworthiness' rather than `value' or `utility' to avoid two possible confusions: (i) `Value' suggest an evaluative rather than a normative property of options. (ii) `Utility' is often understood as a measure of preference satisfaction, while I wish to remain neutral on whether or to what extent an agent's reasons depend on her preferences.}

Any expectational decision theory must assume that degrees of choiceworthiness 
can be represented on an interval scale (i.e., can be given a real-valued representation that is unique up to positive affine transformation), and I will adopt this assumption as well (except briefly in \S \ref{section-stochasticdominancedecisiontheory}). Although stochastic dominance reasoning depends only on 
ordinal choiceworthiness relations, the main line of argument I advance below assumes that choiceworthiness is amenable to a certain kind of cardinal representation (as I will explain in \S \ref{section-formalSetup}). I remain neutral, though, on whether cardinal choiceworthiness 
should be understood as primitive or as a representation of an underlying ordinal relation.



\subsection{Two kinds of expectationalism}

I will understand expectationalism as the following thesis.

\begin{description}
	\item[Expectationalism] An option $O$ is rationally permissible in choice situation $S$ if and only if no option in $S$ has greater expected 
	choiceworthiness.
\end{description}


I formally define expected choiceworthiness in \S \ref{section-formalSetup}. But for now, an option's expected choiceworthiness is the probability-weighted sum (or, in continuous contexts, the probability-weighted integral) of its possible degrees of choiceworthiness, which I will treat as well defined only when it converges absolutely (i.e., independent of the order in which the terms are summed or the limits in the improper integral are taken) or diverges absolutely to $+/- \infty$.\footnote{It is worth noting that I am making some choices in my characterization of expectationalism that might be contested. Expectationalists might hold that possible payoffs (sometimes or always) have a privileged ordering such that they generate a valid expectation even when their probability-weighted sum is only conditionally convergent. They might also deny that unconditional divergence to $+/- \infty$ generates infinite expectations. And they might adopt a different choice rule than the one I have stipulated, viz., that options are permissible iff no alternative has greater expected choiceworthiness. These three assumptions will each play some role in my analysis in \S \ref{section-stochasticdominancedecisiontheory}, but I don't think that modifying any of them would substantially affect my diagnosis of the defects of expectationalism. In each case, I have tried to characterize expectationalism in the way that seems most natural and reasonable.}

It will be important for us to distinguish two versions of expectationalism. One view, which I will call \textit{primitive expectationalism}, holds that cardinal degrees of choiceworthiness are specified independently of 
any ranking of prospects or options under uncertainty---e.g., by purely ethical criteria.\footnote{For defense of this `cardinalist' approach, see for instance \cite{ng1997case}. For one illustration of how cardinal values can be specified independent of a ranking of prospects, see \cite{skyrms2019measuring}.} 
Primitive expectationalism then holds that agents should maximize the expectation of these independently specified values. Another view, which I will call \textit{axiomatic expectationalism}, holds that cardinal choiceworthiness is simply a representation of some ranking of options under uncertainty---e.g., an agent's preference ordering. This ranking is required to satisfy a set of axioms which guarantee that it can be represented as maximizing the expectation of \textit{some} assignment of cardinal values to outcomes or options under certainty. 



\subsection{Arguments for expectationalism}

There are two standard arguments for expectationalism, corresponding to primitive and axiomatic expectationalism respectively: \textit{long-run arguments} and \textit{representation theorems}. 

Long-run arguments invoke the law of large numbers which implies that, as the length of a series of probabilistically independent risky choices goes to infinity, the probability that an expectation-maximizing decision rule will outperform any given alternative converges to certainty \citep{feller1968introduction}. If successful, long-run arguments justify a version of primitive expectationalism: Their conclusion is that the agent should maximize the expectation of a cardinal choiceworthiness function whose values do not represent or depend on the agent's antecedently specified preferences toward risky prospects. 
There is an extensive literature on long-run arguments, but the general consensus is that they are unsuccessful.\footnote{For recent critical treatments, see \citeauthor{buchak2013risk} (\citeyear{buchak2013risk}, pp.\@ 212--8) and \citeauthor{easwaran2014decision} (\citeyear{easwaran2014decision}, pp.\@ 3--4).} 
Among other objections, it's unclear what force long-run arguments have for agents who don't in fact face the relevant sort of long run. And since the standard long-run arguments presuppose an \textit{infinitely} long run of independent gambles, it's therefore unclear what force they have for any actual agent, who will face only a finite series of choices in her lifetime.

Thus, the standard defense of expectationalism in contemporary decision theory appeals instead to \textit{representation theorems}. Representation theorems in decision theory show that, if an agent's preferences satisfy certain putative 
coherence constraints, then there is some assignment of cardinal values to outcomes (a \textit{utility function}) such that the agent can be accurately represented as maximizing its expectation. The two best-known such theorems are due to \cite{vonneumann1947theory} and \cite{savage1954foundations}. The axioms that figure in these theorems are subject to ongoing debate, but the axiomatic approach nevertheless retains the status of decision-theoretic orthodoxy.\footnote{For a survey of axiomatic approaches and objections to the standard axioms, see \cite{briggs2017normative}. For criticism of the axiomatic approach more generally, see \cite{meacham2011representation}.}



\subsection{Expectationalism and risk attitudes toward objective value}

My main interest in this paper is in what risk attitudes we should should adopt toward objective goods that have some natural cardinal structure---e.g., lives saved or lost. And the two versions of expectationalism have very different things to say about this question. Primitive expectationalism implies that, insofar as an option's choiceworthiness increases linearly with the quantity of objective value it produces, we should be exactly risk-neutral toward 
objective goods. But axiomatic expectationalism and the representation theorems that are its foundation do not have this implication.

For instance, suppose you are in a situation where many lives are at risk.  
Suppose that (i) the only thing you care about in this situation is saving lives, (ii) you always prefer saving more lives to saving fewer, and (iii) you value all the lives at stake equally, in the sense that all else being equal, you are indifferent between saving any two lives. 
But you do not yet know how to compare risky prospects. If you accept primitive expectationalism, it is natural to suppose that the choiceworthiness of your options 
is linear in lives saved (though this is not a \textit{logical} consequence of (i)--(iii)), in which case primitive expectationalism implies that you should simply maximize the expected number of lives saved---in other words, you should be \textit{risk-neutral} with respect to lives saved.

But suppose instead you merely believe that you should rank prospects in a way that satisfies, say, the von Neumann-Morgenstern (VNM) axioms. Even given (i)--(iii), and even assuming that the objective value of saving \textit{n} lives increases linearly with \textit{n}, the VNM axioms do not imply that you should maximize expected lives saved. Rather, they merely imply that you should maximize the expectation of \textit{some increasing function} of lives saved. 
This function can be arbitrarily concave or convex, meaning that you can be arbitrarily risk-averse or risk-seeking with respect to lives.\footnote{I am here referring to what are sometimes called `actuarial' risk attitudes, as opposed to the sort of risk attitudes that figure in generalized expected utility theories like Buchak's (\citeyear{buchak2013risk}) REU.} More generally, given any antecedently specified ranking or assignment of cardinal values to options under certainty, VNM and 
other standard axiom systems merely imply that you should maximize the expectation of some increasing function of that ranking or assignment.

This permissiveness has its advantages. For instance, consider the `Pascalian' conclusion imputed to expectationalism in \S \ref{section-intro} that, if there is even a one percent chance of a future in which Earth-originating civilization supports $10^{52}$ happy lives, then the `the expected value of reducing existential risk by a mere one billionth of one percentage point is worth a hundred billion times as much as a billion human lives' \citep[p.\@ 19]{bostrom2013existential}. Primitive expectationalism supports this kind of reasoning. 
Axiomatic expectationalism, on the other hand, can disclaim this reasoning and the seemingly-fanatical conclusions it entails---but only because it places no constraints at all on our risk attitudes toward 
goods like happy lives. 
And in more ordinary cases, this looks like a drawback. For instance, axiomatic expectationalism cannot tell you that you should save $10$ lives with probability $0.5$ 
rather than one life for sure. Pushing the point to more counterintuitive extremes, it cannot tell you that you should save $1000$ lives with probability $0.5$ rather than $10$ lives with probability $0.51$; nor that you should save $1000$ lives for sure rather than $1001$ lives with probability $0.01$. 




Is this a defect in standard axiomatic decision theory? It's not obvious. Some decision theorists will say that it is not the job of decision theory to tell you what your risk attitudes should be toward objective goods like lives saved---rather, that's a job for ethics, or some other branch of normative philosophy. But it's pretty clearly a job for \textit{someone}, wherever we place it on the disciplinary org chart: 
The complete normative theory of choice under uncertainty should tell us that, in a situation where all that matters is saving lives and all the lives at stake have equal 
value, we should prefer to save $1000$ lives with probability $0.5$ rather than $10$ lives with probability $0.51$. So even if these questions are beyond its intended remit, axiomatic expectationalism seems to be \textit{incomplete} as a normative theory of decision-making under uncertainty.

In summary, there are two problems for expectationalism that I am hoping to remedy. First, neither version of expectationalism offers a compelling justification for choosing the option that maximizes the expectation of objective values in ordinary cases 
where it seems clear that this is what we should do. Primitive expectationalism relies on the dubious appeal to hypothetical long runs, while axiomatic expectationalism does not attempt to justify this conclusion in the first place. Second, insofar as expectationalism \textit{does} offer a justification for maximizing expected objective value, it goes too far, committing us to Pascalian fanaticism in cases involving minuscule probabilities of astronomical payoffs.\footnote{This is true of primitive expectationalism, and also of the most natural strategy for placing constraints on risk attitudes toward objective value within the axiomatic framework
---namely, an appeal to `aggregation theorems' like that of \cite{harsanyi1955cardinal}.} I aim both to provide a stronger justification for choosing options that maximize expected objective value in ordinary cases, and in so doing to draw a principled line between those ordinary cases and extreme, Pascalian cases.

It is important to note, however, that the arguments I advance below will interact very differently with primitive and axiomatic expectationalism.  
Specifically: I will propose that 
stochastic dominance can provide a sufficient criterion of rational choice under uncertainty. This view is a rival to both primitive and axiomatic expectationalism. The primary 
motivation for this view will be the results in \S \ref{section-results}. And while the primitive expectationalist cannot take any advantage of these results, the axiomatic expectationalist can: As we will see in \S \ref{section-addOnToExpectationalism}, those who accept the standard axioms can interpret these results 
as furnishing a friendly `add-on' to standard axiomatic decision theory. The main advantages of my proposed view over axiomatic expectationalism will be that it can recover strong practical conclusions about choice under uncertainty from something much weaker and less controversial than the standard axiom systems, and that it better handles the range of problem cases surveyed in \S \ref{section-intro}.

\section{Stochastic dominance}
\label{section-stochasticDominance}





Option $O$ \textit{first-order stochastically dominates} option $P$ 
if and only if

\begin{enumerate}
	
	\item For any payoff \textit{x}, the probability that $O$ yields a payoff at least as good as \textit{x} is equal to or greater than the probability that $P$ yields a payoff at least as good as \textit{x}, and
	
	\item For some payoff \textit{x}, the probability that $O$ yields a payoff at least as good as \textit{x} is strictly greater than the probability that $P$ yields a payoff at least as good as \textit{x}.
	
\end{enumerate}

There are also second- and higher-order stochastic dominance relations, which are less demanding than first-order stochastic dominance. (For a survey, 
see Ch.\@ 3 of  \citeauthor{levy2016stochastic} (\citeyear{levy2016stochastic}).) But since we will only be concerned with the first-order relation, I 
omit the qualifier and use `stochastic dominance' to mean `\textit{first-order} stochastic dominance'.


Stochastic dominance is a generalization of the familiar \textit{statewise} dominance relation that holds between \textit{O} and \textit{P} whenever \textit{O} yields at least as good a payoff as \textit{P} in every possible state, and a strictly better payoff in some state. To illustrate: Suppose that I am going to flip a fair coin, and I offer you a choice of two tickets. The Heads ticket will pay \$1 for heads and nothing for tails, while the Tails ticket will pay \$2 for tails and nothing for heads. 
The Tails ticket does not \textit{statewise} dominate the Heads ticket because, if the coin lands Heads, the Heads ticket yields a better payoff. But the Tails ticket does \textit{stochastically} dominate the Heads ticket. There are three possible payoffs: 
winning \$0, winning \$1, and winning \$2. The two tickets offer the same probability of a payoff at least as good as \$0, namely 1. And they offer the same probability of an payoff at least as good as \$1, namely 0.5. But the Tails ticket offers a greater probability of a payoff at least as good as \$2, namely 0.5 rather than 0.

Stochastic dominance is generally seen as giving a necessary condition for rational choice:

\begin{description}
	\item[Stochastic Dominance Requirement (SDR)] An option $O$ is rationally permissible in situation $S$ only if it is not stochastically dominated by any other option in $S$.
\end{description}
This principle is on a strong \textit{a priori} footing. 
Various formal arguments can be made in its favor. For instance, if $O$ stochastically dominates $P$, then $O$ can be made to \textit{statewise} dominate $P$ by an appropriate permutation of equiprobable states in a sufficiently fine-grained partition of the state space \citep{easwaran2014decision,bader2018stochastic}. 
So if one is rationally required to reject statewise dominated options, and if the rational permissibility of an option depends only on its prospect and not on which payoffs are associated with which states, then one is rationally required to reject stochastically dominated options as well. The claim that an option's rational permissibility depends only on its prospect reflects the idea that all normatively significant features of an outcome are captured by the payoff value assigned to that outcome, so that as a conceptual matter an agent must be indifferent between receiving a given payoff in one state or another. If, say, you prefer winning \$0 with a Heads ticket to winning \$0 with a Tails ticket, 
then this should be reflected in the values assigned to the payoffs, in which case the Tails ticket would no longer stochastically dominate the Heads ticket.




More informally, it is unclear how one could ever \textit{reason} one's way to choosing a stochastically dominated option $P$ over the option $O$ that dominates it. For any feature of $P$ that one might point to as grounds for choosing it, there is a persuasive reply: However choiceworthy $P$ might be in virtue of possessing that feature, $O$ is equally or more likely to be at least that choiceworthy. And conversely, for any feature of $O$ one might point to as grounds for rejecting it, there is a persuasive reply: However unchoiceworthy $O$ might be in virtue of possessing that feature, $P$ is equally or more likely to be at least that unchoiceworthy. To say that $O$ stochastically dominates $P$ is in effect to say that there is no feature of $P$ that can provide a \textit{comparative} 
justification for choosing it over $O$.

For reasons like these, SDR is almost entirely uncontroversial in normative decision theory. In particular, it is much less controversial than the 
axioms of expected utility theory: The most widely discussed alternatives to 
axiomatic expectationalism, which give up one or more of those axioms  (e.g., rank-dependent expected utility (RDU) \citep{quiggin1982theory} and its philosophical cousin, risk-weighted expected utility (REU) \citep{buchak2013risk}) all satisfy stochastic dominance.\footnote{SDR \textit{has} been challenged in certain special contexts---specifically, in the context of gambles without finite expectations by \cite{seidenfeld2009preference} and \cite{lauwersvallentyne2016decision}, and in the context of incomparability/incompleteness by \cite{bales2014decision} and \cite{schoenfield2014decision}. I will briefly discuss these challenges, and explain why I find them unpersuasive, in notes \ref{footnote-lauwersVallentyneObjection} and \ref{footnote-opaqueSweeteningObjection} respectively.


In descriptive decision theory, the original version of prospect theory allowed stochastic dominance violations, and largely for that reason was superseded by \textit{cumulative} prospect theory \citep{tversky1992advances}, which satisfies stochastic dominance.} 



My aim, however, is to defend stochastic dominance as not just a necessary but also a \textit{sufficient} criterion for rational permissibility. Let's call this the \textit{stochastic dominance theory of rational choice}.

\begin{description}
	\item[Stochastic Dominance Theory of Rational Choice (SDTR)] An option $O$ is rationally permissible in situation $S$ if and only if it is not stochastically dominated by any other option in $S$.
\end{description}


Whereas SDR is about as uncontroversial as normative principles come, SDTR is radically revisionary. The only previous advocate of this view that I am aware of is \cite{manski2011actualist} (who defends it on grounds very different from those I will give below, but with which I am broadly sympathetic).\footnote{Manski is somewhat equivocal between SDTR and the even more revisionary view that an option is rationally permissible iff it is not \textit{(weakly) statewise} dominated. He seems to think that choosing a stochastically dominated option merits \textit{some} form of negative normative appraisal, while being reluctant to apply the epithet `irrational'.}

What is the relationship between SDTR and expectationalism? In a broad range of cases (in particular, whenever the expected choiceworthiness of all options is finite---i.e., neither infinite nor undefined), $O$ stochastically dominates $P$ only if it has greater expected choiceworthiness.
So in these cases, SDTR is strictly more permissive than expectationalism. But as we will see in \S \ref{section-stochasticdominancedecisiontheory}, there are other cases where SDTR can deliver guidance that expectational reasoning cannot, and is therefore less permissive.

Like axiomatic expectationalism, SDTR does not constrain an agent's risk attitudes toward objective goods (in the absence of background uncertainty): 
In a situation where all that matters is saving lives, saving more lives is always better than saving fewer, and all the lives at stake have equal value, stochastic dominance does not require you to save $10$ lives with probability $0.5$ rather than one life for sure, or even to save $1000$ lives with probability $0.5$ rather than $10$ lives with probability $0.51$.\footnote{In fact, in this sort of case, SDTR and axiomatic expectationalism are very closely related: Given a fixed ordering of payoffs, 
it is possible to prefer $O$ to $P$ while satisfying the VNM axioms iff $P$ does not stochastically dominate $O$ (i.e., iff SDTR permits you to choose $O$ over $P$). The difference is that axiomatic expectationalism imposes `global' constraints on an agent's preferences (e.g., Independence and Continuity) that SDTR does not.} This means that \textit{primitive} expectationalism has an apparent advantage over both axiomatic expectationalism and SDTR: It can explain why, in ordinary cases, you ought to maximize the expectation of objective goods like lives saved.

But, I will argue, this advantage is only apparent: 
Under realistic levels of background uncertainty, SDTR can effectively constrain an agent's risk attitudes toward objective goods, 
recovering many of the attractive 
implications of primitive expectationalism---while still avoiding its fanatical implications in Pascalian cases. 
In \S \ref{section-results}, we will see how this can happen. But first, we must introduce a formal framework for describing decisions under background uncertainty.

\section{Formal setup}
\label{section-formalSetup}



A \textit{choice situation} is an ordered triple $S = \langle A, \textrm{\textbf{O}}, \bupdf \rangle$, where $A$ is an agent, \textbf{O} is a 
set of \textit{options} $\{O_1, O_2, ...,  O_m\}$, and $\bupdf$ is a probability density function (PDF) over the real numbers that represents the agent's \textit{background uncertainty} in \textit{S}. We identify each option $O_i \in \textrm{\textbf{O}}$ with its \textit{simple prospect}, a finite 
set of ordered pairs $O_i = \{ \langle v_1^i,p_1^i \rangle, \langle v_2^i,p_2^i \rangle, ..., \langle v_n^i,p_n^i \rangle \}$, 
where $v_j^i \in \mathbb{R}$ is a possible \textit{simple payoff} and $p_j^i \in (0,1]$ is the probability of obtaining that simple payoff associated with $O_i$.\footnote{The restriction to options with finitely many simple payoffs is a useful simplifying assumption for the discussion in \S \S \ref{section-results}--\ref{section-twoModestConclusions}. I will relax this assumption, along with other simplifying assumptions in the present framework (e.g., that payoffs are always comparable), to discuss problem cases like the St.\@ Petersburg and Pasadena games in \S \ref{section-stochasticdominancedecisiontheory}.} 
I will generally omit the superscripts on payoffs and probabilities, where there is no risk of confusion. The $p_j$ are all positive (i.e., we ignore simple payoffs with probability 0) and sum to 1.

I remain neutral on the interpretation of these probabilities, in two ways. First, I leave it unspecified whether $p_j^i$ represents the conditional probability $\Pr(v_j^i | O_i)$ or the causal probability 
$\Pr(v_j^i \, \backslash \, O_i)$ \citep[pp.\@ 161ff]{joyce1999foundations}, and hence remain neutral between evidential and causal decision theory. Second, I leave it unspecified whether these probabilities are subjective or epistemic. 



Intuitively, the simple payoff of an option is what the option itself yields. 
Crucially, however, an option's \textit{overall} payoff depends not just on its simple payoff, but also on what I will call a \textit{background payoff}. A background payoff is, roughly, \textit{what the agent starts off with}, or \textit{the component of the overall outcome/payoff that does not depend on her choice}. 
As a mundane illustration: Suppose that a young person is deciding how to invest some money in her retirement account, and that 
her only concern in this context is her net worth when she retires. 
Her options are various funds 
she can 
invest in. The simple payoff of buying some shares in fund $F_i$ (call this option $O_i$) is the value those shares will have when she retires. But the \textit{overall} payoff of $O_i$---the thing she ultimately cares about---is her total net worth at retirement, if she now invests in $F_i$. This overall payoff is the sum of her simple payoff (the future value of her $F_i$ shares) plus a background payoff (the value of all her other assets).


Just as an agent may be uncertain about an option's simple payoff, 
she may 
be uncertain about her background payoff. 
This is what I will call \textit{background uncertainty}. The defining feature of 
background uncertainty is its independence from other features of the choice situation. In particular, $A$'s background payoff in $S$ is probabilistically independent of (i) which option she chooses and (ii) which simple payoff she receives from her chosen option. 
Thus, $A$'s background uncertainty captures uncertainties that apply to \textit{all} the options in situation \textit{S}, rather than uncertainties about any one option in particular. We will describe $A$'s background uncertainty 
by means of a continuous random variable---her \textit{background prospect}---with probability density function $\bupdf$, such that the probability of a background payoff in the interval $[n,m]$ 
is given by 
$\int_{n}^{m} \bupdf(x) \, dx$. (Again, these probabilities can be interpreted as either conditional or causal, and as either subjective or epistemic.)




I have already mentioned one possible source of background uncertainty (concerning financial decisions), 
but my primary focus 
will be on a different source: I will assume that agents should assign at least some normative weight to \textit{aggregative consequentialist} considerations, i.e., they should measure the choiceworthiness of an option \textit{at least in part} by the total amount of value in the resulting world. 
Such agents will 
be in a state of background uncertainty because they are 
uncertain how much value there is in the world \textit{to begin with}, independent of their present choice. 
In this case, we can understand $\bupdf$ as giving the probability that, excluding the outcome of $A$'s present choice, the world contains value equivalent to between \textit{n} and \textit{m} units of choiceworthiness, via 
$\int_{n}^{m} \bupdf(x) \, dx$.\ifanon\else\footnote{Some definitions: \textit{Aggregative consequentialist} ethical theories 
assert that the choiceworthiness of an option is entirely determined by the overall value of the resulting world, and the overall value of a world is measured by an impartial, additively separable axiology. \textit{Additive separability} means that the axiology can be represented as ranking worlds by the sum of degrees of value and disvalue realized by each value-bearing entity (e.g., 
welfare subject) in that world. \textit{Impartiality} means that this sum does not give different weight to otherwise similar value-bearing entities based on spatiotemporal location or (vaguely) other morally irrelevant considerations. An agent, or a normative theory, `gives normative weight to aggregative consequentialist considerations' if she/it regards the overall value of the resulting world, as measured by an impartial, additively separable axiology, as making a \textit{pro tanto} contribution to the choiceworthiness of an option---i.e., supplying \textit{pro tanto} reasons for or against particular options.}\fi

It might seem that background uncertainty has no bearing on what an agent ought to do, since it does not affect the \textit{relative} choiceworthiness of her options. Over \S\S \ref{section-results}--\ref{section-relevanceofbackgrounduncertainty}, however, I will make the case that background uncertainty can have a great deal of practical significance, and so 
must be included in our representation of choice situations. 


The \textit{payoff} of an option 
is simply its overall degree of objective choiceworthiness, as determined by the combination of its simple and background payoffs. 
Specifically, I will assume that an option's payoff can be represented as the \textit{sum} of its constituent simple and background payoffs---i.e., that 
we can assign real numbers to simple and background payoffs such that one overall payoff is at least as good as another just in case 
the sum of the numbers assigned to its constituent simple and background payoffs is at least as great. Call this \textit{additive separability} between simple and background payoffs.

Additive separability is not as strong an assumption as it might sound: In particular, it does not require us to assume that payoffs have any primitive cardinal structure. Suppose there is a set $S$ of possible simple payoffs and a set $B$ of possible background payoffs, and that the set of possible overall payoffs $S \times B$ is totally preordered by a relation $\succcurlyeq_p$. Then additive separability amounts to the assumption that $\langle S, B, \succcurlyeq_p \rangle$ forms an \textit{additive conjoint structure}. This involves satisfying a number of purely ordinal axioms, the most important of which is an ordinal separability condition to the effect that, if we know that two overall payoffs $p_i$ and $p_j$ have one component in common (i.e., involve the same simple background payoff or the same background payoff), we can learn whether $p_i \succcurlyeq_p p_j$ by learning the distinctive component of each payoff.\footnote{The other axioms that characterize additive conjoint structures are mainly technical---e.g. (i) requiring that the sets $S$ and $B$ are sufficiently rich that for any $s_i, s_j \in S$, $b_k \in B$ there is a $b_l \in B$ such that $\langle s_i, b_k \rangle \sim_p \langle s_j, b_l \rangle$, and (ii) requiring that no payoff is infinitely better than another, in the sense that we can always `get from' one payoff to another by repeatedly substituting a more preferred component for a less preferred component (e.g., repeatedly substituting $s_i$ for $s_j$, where $\forall b \in B (\langle s_i,b \rangle \succ_p \langle s_j, b \rangle)$, to create an ascending series of overall payoffs), in a finite number of steps. For a full characterization of additive conjoint structures and a proof that all such structures have an additively separable representation, see \citeauthor{krantz1971foundations} (\citeyear{krantz1971foundations}, pp.\@ 245-266).} 
If an additively separable representation of payoffs exists, then it is 
unique up to choice of zero elements in $S$ and $B$ and a unit element in \textit{either} $S$ or $B$. 
Thus, the real numbers used to designate simple, background, and overall payoffs can be either taken as given or understood to represent an underlying ordinal relation on ordered pairs of simple and background payoffs.


The \textit{prospect} of $O_i$ is the probability distribution it yields over payoffs. 
Given the assumptions of independence and additive separability, we can express prospects as follows: Where $O_i = \{ \langle v_1,p_1 \rangle, \langle v_2,p_2 \rangle, ..., \langle v_n,p_n \rangle\}$ and $A$'s background uncertainty is described by $\bupdf$, the \textit{prospect} of $O_i$ is described by $\bupdf_i(x) = p_1\bupdf(x-v_1) + p_2\bupdf(x-v_2) + ... + p_n\bupdf(x-v_n)$. Formally, $\bupdf_i$ is a \textit{mixture distribution}, a convex combination of $n$ copies of the background prospect $\bupdf$, each corresponding to a possible simple payoff $\langle v_i,p_i \rangle$, and therefore translated along the $x$ axis by the value of that simple payoff ($v_i$) and weighted by the probability of receiving that simple payoff ($p_i$). Since the $p_i$ sum to 1, $\bupdf_i$ is a probability density function.

It will sometimes be useful to represent a prospect by its \textit{cumulative distribution function} (CDF), denoted $\bucdf_i(x) = \int_{- \infty}^{x} \bupdf_i(y) \, dy$, which gives the probability of the prospect yielding a payoff less than or equal to $x$. Even more useful for visualizing stochastic dominance relations is the \textit{complementary} cumulative distribution function (CCDF), 
$\ccdf{\bucdf_i}(x) = 1 - \bucdf_i(x)$, which 
gives the probability of a payoff $\geq x$.

We can now formally define 
expected choiceworthiness and stochastic dominance, which are respectively a property of and a relation on 
(overall) prospects. 
The expected choiceworthiness of option $O_i$ is given by $\mathbb{E}(O_i) = \int_{- \infty}^{\infty} x \bupdf_i(x) \, dx = \lim\limits_{b \to \infty} \int_{0}^{b} x \bupdf_i(x) \, dx + \lim\limits_{a \to - \infty} \int_{a}^{0} x \bupdf_i(x) \, dx$ (with these limits allowed to take infinite values). And stochastic dominance between options $O_i$ and $O_j$ can be expressed as $\sdArgs{O_i}{O_j} \Leftrightarrow \forall x (\bucdf_j(x) \geq \bucdf_i(x)) \land \exists x (\bucdf_j(x) > \bucdf_i(x))$---or equivalently, $\forall x (\ccdf{\bucdf_i}(x) \geq \ccdf{\bucdf_j}(x)) \land \exists x (\ccdf{\bucdf_i}(x) > \ccdf{\bucdf_j}(x))$.

\section{Stochastic dominance under background uncertainty}
\label{section-results}

This section describes the general phenomenon of background uncertainty generating stochastic dominance and states two central results, establishing respectively a sufficient and a necessary condition for stochastic dominance under background uncertainty. The first result shows that, if $O$'s simple prospect is expectationally superior to $P$'s, then under sufficient background uncertainty, $O$ will stochastically dominate $P$. The second result shows that, when 
the balance of expectations depends on minuscule probabilities of astronomical payoffs, much greater background uncertainty is needed to generate stochastic dominance, so that SDTR is more permissive in more Pascalian choice situations. 

For background uncertainty to generate stochastic dominance means that, for some options $O$ and $P$, $O$'s prospect stochastically dominates $P$'s as a result of the agent's background uncertainty, even though $O$'s \textit{simple} prospect does not stochastically dominate $P$'s.\footnote{Interestingly, despite a very large literature on stochastic dominance, the possibility of background uncertainty generating stochastic dominance appears to have gone unremarked until quite recently, when it was was noticed independently by myself and \cite{pomatto2020stochastic}. To my knowledge, it has not been noted or discussed elsewhere. Pomatto et al's interests are somewhat different from mine, and our main results are non-overlapping.}  
The crucial condition under which this phenomenon can become widespread---and therefore, the condition under which the results below become interesting---is that the agent's background prospect has \textit{exponential or heavier tails}, 
meaning that it is bounded below in the tails by some member of the \textit{Laplace} (or \textit{double-exponential}) family of distributions. Laplace distributions have PDFs of the form $L(x|\mu,\rho) = \frac{1}{2 \rho} e^{- \frac{|x - \mu|}{\rho}}$, where $\mu$ is a \textit{location parameter} that determines where the distribution is centered, and $\rho$ is a \textit{scale parameter} that determines how `spread out' it is. To say that $\bupdf$ is bounded below in the tails by a Laplace distribution means that there are some real numbers $\mu$, $\rho$, and $c$ such that, if $|x| > c$, then $\bupdf(x) \geq L(x|\mu,\rho)$.

More precisely, it will be convenient to focus on a slightly stronger condition, namely, that the \textit{decay rate} $\frac{|\bupdf'(x)|}{\bupdf(x)}$ of the agent's background prospect has a finite upper bound. I will say that a $\bupdf$ satisfying this condition has \textit{large tails}. This is a slight abuse of terminology, since the preceding condition is not strictly a `tails' condition. But it is, in practice, nearly extensionally equivalent to the `exponential or heavier tails' condition defined above---in particular, all common parameterized families of probability distributions satisfy one condition if and only if they satisfy the other.\footnote{Large tails are not a necessary condition for background uncertainty to generate stochastic dominance. (In particular, local violations of the large tails condition, e.g.\@ by a vertical asymptote in $\bupdf$, do not always substantially weaken the stochastic dominance constraints that $\bupdf$ imposes.) But it is a very good approximate criterion (as far as I have been able to discover, anyway) for the circumstances in which stochastic dominance can strongly constrain risk attitudes toward simple prospects, and as we will see, it has an important connection with the sufficient condition for stochastic dominance identified by the Sufficiency Theorem below, making it a natural condition on which to focus.}

While any large-tailed background prospect will generate 
\textit{some} new stochastic dominance relations among options, the strength of the resulting stochastic dominance constraints 
depends on the \textit{dispersion} of $\bupdf$. Intuitively, dispersion describes how `spread out' a distribution is. There are many ways of measuring dispersion, but a simple measure 
is \textit{interquartile range} ($\IQR$), the distance between a distribution's 25th and 75th percentiles (i.e., the width of its 50\% confidence interval). 
We can change the dispersion of a background prospect $\bupdf$ 
by applying a \textit{rescaling}, 
transforming it to $\bupdf_s(x) = \frac{1}{s} \bupdf(\frac{x - a}{s})$ for some constants $s > 0$ and $a$. This increases (for $s > 1$) or decreases (for $s < 1$) the dispersion by a factor of $s$. An increasing rescaling `stretches' $\bupdf$ horizontally, 
while otherwise preserving 
its shape.\footnote{For distributions in a parameterized family, like Laplace distributions, we can achieve the same effect by increasing the scale parameter.} 
As we will see, given a large-tailed $\bupdf$, stochastic dominance approximates the ranking of options by the expectations of their simple prospects ever more closely under increasing rescalings of $\bupdf$. (As we will see in \S \ref{section-illustrationsAndImplications}, though, the dispersion of $\bupdf$ need not be particularly large to generate fairly strong constraints.) 
\ifanon \else In \S \ref{section-sourcesofbackgrounduncertainty}, I will argue that a large-tailed $\bupdf$ with high dispersion is rationally warranted 
for agents with ordinary evidence who assign normative weight to aggregative consequentialist considerations. For now, I take it for granted.\fi 




I begin in \S \ref{section-intuitivedescription} with an intuitive explanation 
of how background uncertainty generates stochastic dominance.
In \S\S \ref{section-mainresult}--\ref{section-necessitythm}, I state and describe the two main results. %
In \S \ref{section-illustrationsAndImplications}, I draw out their practical implications, describing both how tightly SDTR can constrain our risk attitudes toward ordinary gambles in the presence of moderate background uncertainty, and how much looser those constraints become for more Pascalian choices.

\subsection{How background uncertainty generates stochastic dominance}
\label{section-intuitivedescription}

Suppose you face a risky option that will either save two lives (with probability 0.5) or cause one death (with probability 0.5). Suppose that the lives at stake all have equal 
value and 
there are no other normatively relevant considerations (e.g., deontological constraints) that should influence your choice besides maximizing the  number of lives saved. Call this option the \textit{Basic Gamble}.

\newcommand*\BasicGamble{G}

\begin{description}
	\item[ ] \textbf{Basic Gamble (\textit{\BasicGamble})} $\{ \langle -1, 0.5 \rangle, \langle 2, 0.5 \rangle \}$
\end{description}
Suppose that your only other option is what we will call the \textit{Null Option}.

\newcommand*\NullOption{N}

\begin{description}
	\item[ ] \textbf{Null Option (\textit{\NullOption})} $\{ \langle 0, 1 \rangle \}$
\end{description}
Intuitively, the Null Option can be thought of as the option of `doing nothing', and simply accepting your background payoff as your overall payoff.

\begin{figure}
	\centering
	\includegraphics[width=0.8\linewidth]{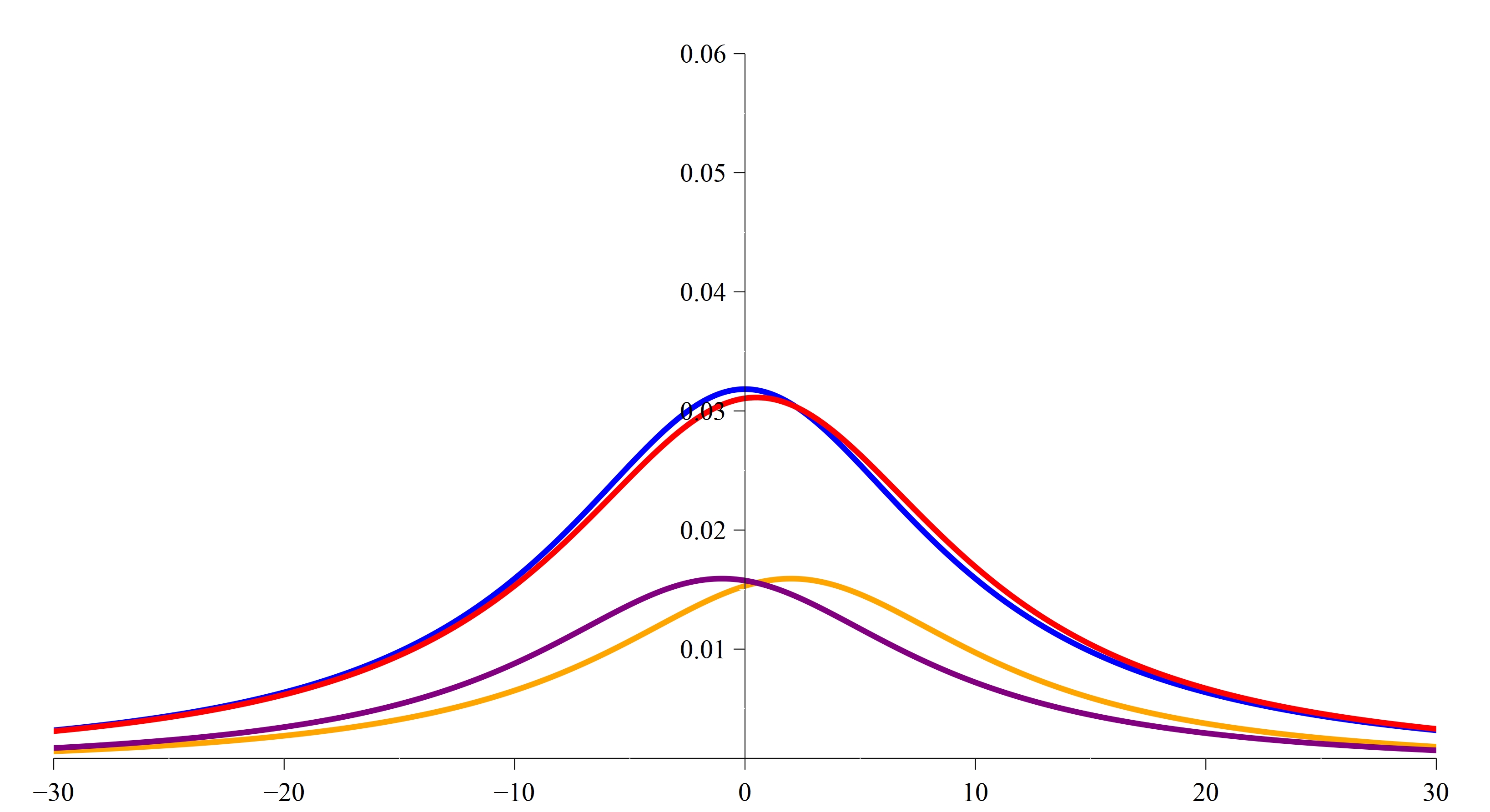}
	\caption[Basic Gamble: PDFs]{PDFs representing the prospects of the Null Option (blue) and the Basic Gamble (red), given a background prospect described by a Cauchy distribution with a location parameter of 0 and a scale parameter of 10. Purple and orange curves are `half PDFs' representing the two possible outcomes of the Basic Gamble: They are obtained from the background distribution $\bupdf$ by multiplying 
	by $0.5$ (representing the $0.5$ probabilities of each simple payoff), then translating 
	by $+2$ 
	and $-1$ 
	respectively (representing the magnitudes of the simple payoffs). The prospect of the Basic Gamble is then obtained by summing the orange and purple curves. [Blue: $\bupdf_\NullOption(x) = \bupdf(x) = \left( 10 \pi ( 1 + (\frac{x}{10})^2 ) \right)^{-1}$. Purple: $\bupdf^\BasicGamble_1(x) = 0.5\bupdf(x+1)$. Orange: $\bupdf^\BasicGamble_2(x) = 0.5\bupdf(x-2)$. Red: $\bupdf_\BasicGamble(x) = \bupdf^\BasicGamble_1(x) + \bupdf^\BasicGamble_2(x)$.]}
	\label{figure-pdfs}
\end{figure}

In the absence of background uncertainty, neither of these options is stochastically dominant: $\BasicGamble$ gives a greater probability of a payoff $\geq 2$, but $\NullOption$ gives a greater probability of a payoff $\geq 0$. 
But suppose you are in a state of background uncertainty, described by a 
PDF $\bupdf$. $\NullOption$'s prospect, then, is simply given by $\bupdf_\NullOption(x) = \bupdf(x)$. $\BasicGamble$'s prospect is given by $\bupdf_\BasicGamble(x) = 0.5\bupdf(x-2) + 0.5 \bupdf(x + 1)$. Visually, we can think of $\BasicGamble$'s prospect as follows (Fig.\@ \ref{figure-pdfs}): We make two half-sized copies of $\bupdf$, corresponding to the two possible outcomes of $\BasicGamble$, each of which has probability $0.5$. We then translate one of those copies two units to the right (representing a gain of 2, relative to the background payoff) and the other one unit to the left (representing a loss of 1, relative to the background payoff). Finally, we add these two half-PDFs together, obtaining the new PDF $\bupdf_\BasicGamble$.

For each possible payoff $x$, choosing $\BasicGamble$ rather than $\NullOption$ makes both a positive contribution and a negative contribution to the probability of a payoff 
$\geq x$.

\begin{itemize}
	\item Positive contribution: If $\bupdf$ yields a background payoff in the interval $[x - 2, x)$ and $\BasicGamble$ yields a simple payoff of $2$, then $\BasicGamble$ results in a payoff $\geq x$ where $\NullOption$ would have resulted in a payoff $< x$. The probability of a background payoff in the interval $[x - 2, x)$ is given by $\int_{x - 2}^{x} \bupdf(y) \, dy$, and the probability that $\BasicGamble$ yields a simple payoff of $2$ is $0.5$. Since these probabilities are independent, we can multiply them. So the possibility of a positive simple payoff from $\BasicGamble$ increases the probability of an overall payoff $\geq x$ by $0.5 \int_{x - 2}^{x} \bupdf(y) \, dy$.
	
	\item Negative contribution: If $\bupdf$ yields a background payoff in the interval $[x, x + 1)$ and $\BasicGamble$ yields a simple payoff of $-1$, then $\BasicGamble$ results in a payoff $< x$ where $\NullOption$ would have resulted in a payoff $\geq x$. The probability of a background payoff in the interval $[x, x + 1)$ is given by $\int_{x}^{x + 1} \bupdf(y) \, dy$, and the probability that $\BasicGamble$ yields a simple payoff of $-1$ is $0.5$. 
	So the possibility of a negative simple payoff from $\BasicGamble$ decreases the probability of an overall payoff $\geq x$ by $0.5 \int_{x}^{x + 1} \bupdf(y) \, dy$.
	
\end{itemize}

\begin{figure}
	\centering
	\includegraphics[width=0.8\linewidth]{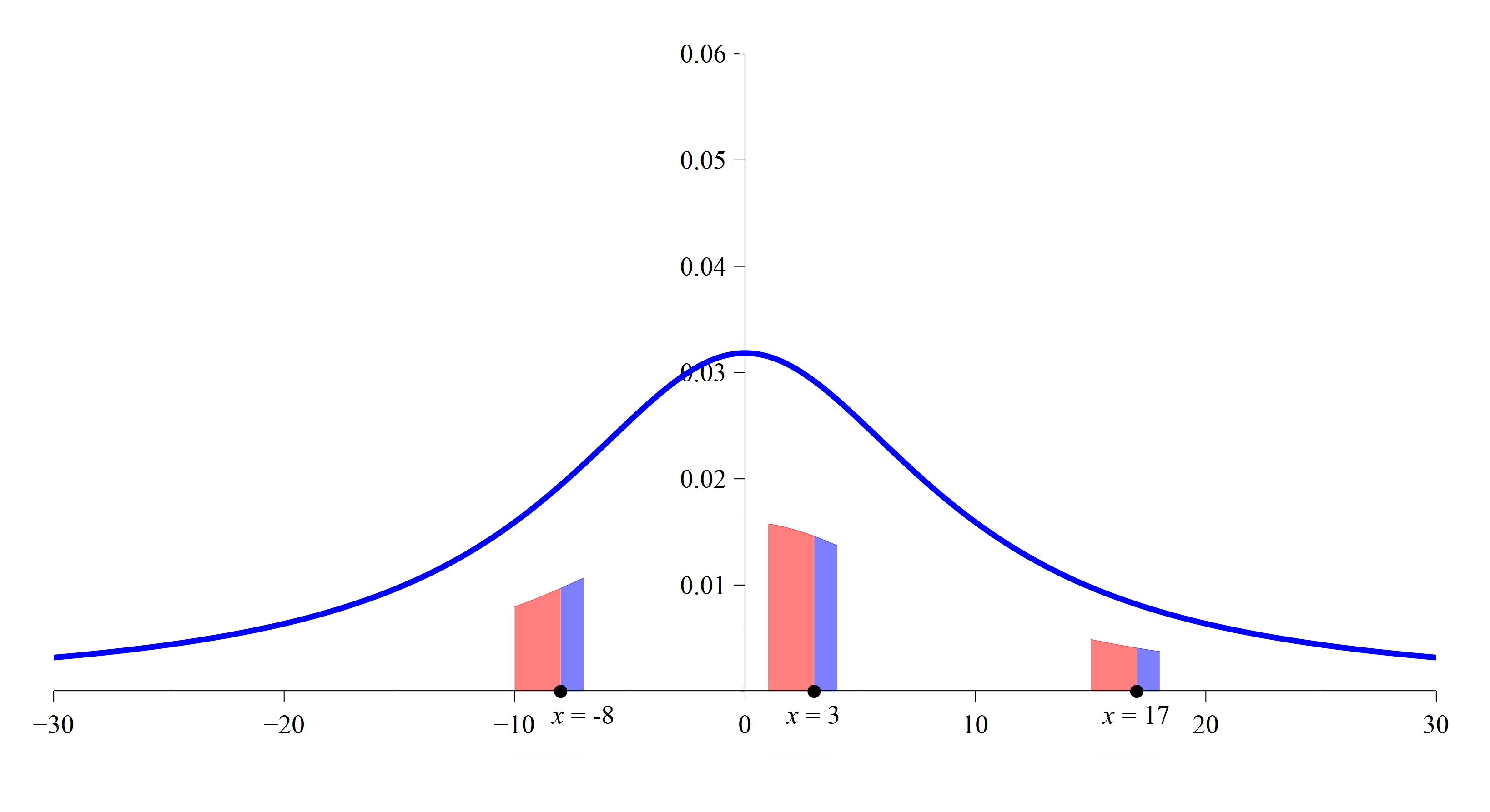}
	\caption{Red areas represent the joint probability of a simple payoff of $2$ and a background payoff in the interval $[x-2,x)$ (for $x = -8$, $3$, $17$). Blue areas represent the joint probability of a simple payoff of $-1$ and a background payoff in the interval $[x,x+1)$. $\BasicGamble$ stochastically dominates $\NullOption$ if for every $x$, $0.5\int_{x-2}^{x} \bupdf(y) \, dy$ (red) is greater than $0.5\int_{x}^{x+1} \bupdf(y) \, dy$ (blue).}
	\label{figure-shadedIncAndDecRegions}
\end{figure}

\begin{figure}
	\centering
	\includegraphics[width=0.8\linewidth]{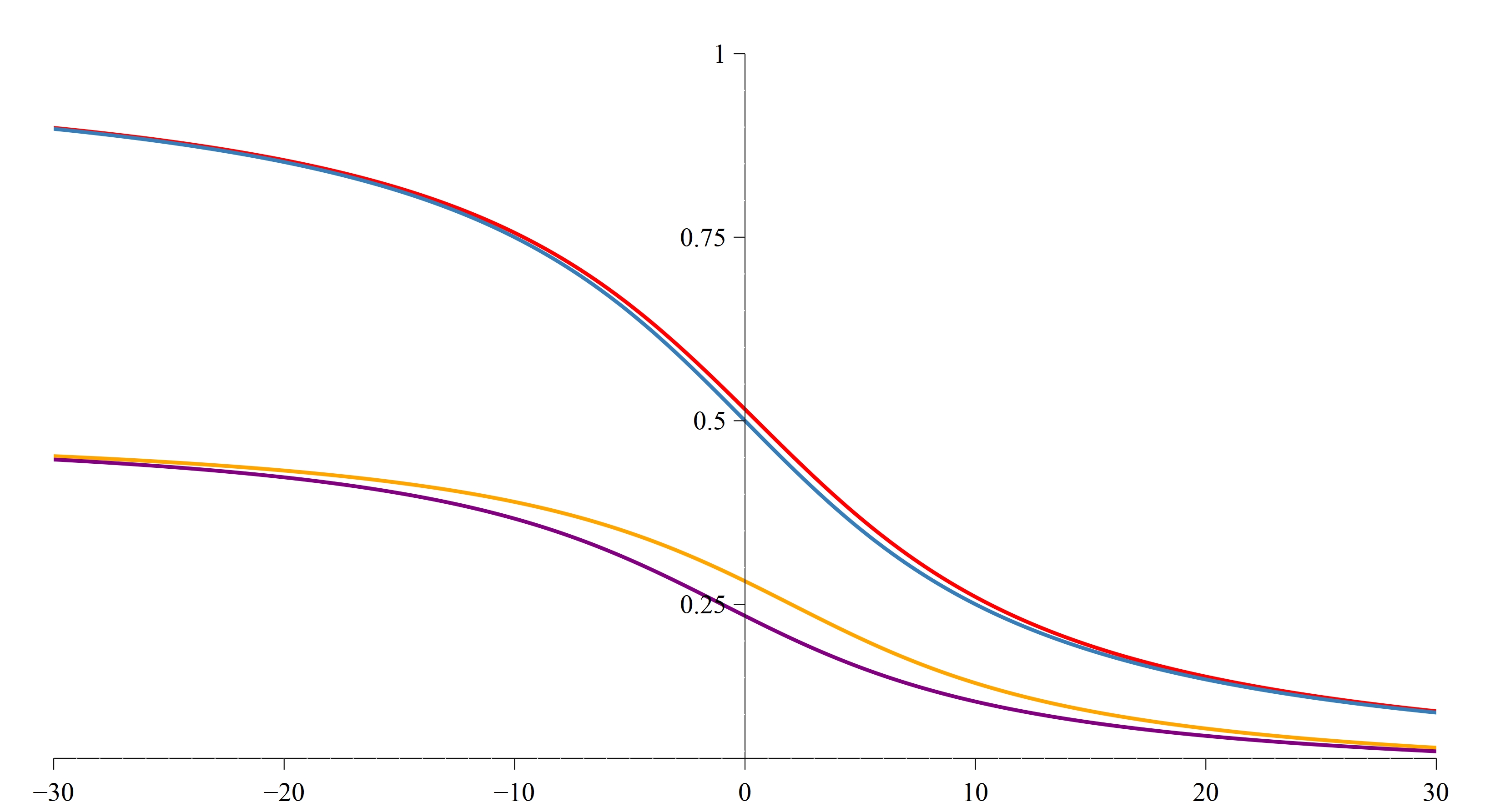}
	\caption[Basic Gamble (CCDFs)]{CCDFs (and `half CCDFs') corresponding to the PDFs (and `half PDFs') in Fig.\@ \ref{figure-pdfs}. The blue curve gives the probability that the Null Option yields a payoff $\geq x$. The red curve gives the probability that the Basic Gamble yields a payoff $\geq x$. Purple and orange curves again represent the two possible simple payoffs of the gamble. $\ccdf{\bucdf}_\BasicGamble$ (red) is everywhere slightly greater than $\ccdf{\bucdf}_\NullOption$ (blue), indicating stochastic dominance. 
	[Blue: $\ccdf{\bucdf}_\NullOption(x) = \ccdf{\bucdf}(x) = \frac{1}{\pi} \tan^{-1}\left( \frac{x}{10} \right) + 0.5$. Purple: $\ccdf{\bucdf}^\BasicGamble_1(x) = 0.5\ccdf{\bucdf}(x+1)$. Orange: $\ccdf{\bucdf}^\BasicGamble_2(x) = 0.5\ccdf{\bucdf}(x-2)$. Red: $\ccdf{\bucdf}_\BasicGamble(x) = \ccdf{\bucdf}^\BasicGamble_1(x) + \ccdf{\bucdf}^\BasicGamble_2(x)$.]}
	\label{figure-ccdfs}
\end{figure}
%

Thus, $\BasicGamble$ offers a greater probability than $\NullOption$ of a payoff $\geq x$ iff $0.5 \int_{x - 2}^{x} \bupdf(y) \, dy > 0.5 \int_{x}^{x + 1} \bupdf(y) \, dy$. 
If this inequality holds for every $x$, then $\BasicGamble$ stochastically dominates $\NullOption$ (see Fig.\@ \ref{figure-shadedIncAndDecRegions}). Formally (and canceling the $0.5$'s):

$$\forall x \left( \int_{x-2}^{x}\bupdf\left(y\right) \, dy > \int_{x}^{x+1}\bupdf(y) \, dy \right) \Rightarrow \sdArgs{\BasicGamble}{\NullOption}$$

If $\bupdf$ is unimodal (i.e., strictly decreasing in either direction away from a central peak), then this condition will be trivially satisfied for values of $x$ in the right tail: Since $\bupdf$ is decreasing in the right tail, $\int_{x - 2}^{x} \bupdf(y) \, dy$ will clearly be greater than $\int_{x}^{x + 1} \bupdf(y) \, dy$, being both `wider' and `taller'. The interesting question is whether it holds in the left tail. A sufficient condition for it to do so is that the value of $\bupdf$ never decreases by more than a factor of 2 in an interval of length 3: In this case, $\int_{x - 2}^{x} \bupdf(y) \, dy$ is everywhere greater than $\int_{x}^{x + 1} \bupdf(y) \, dy$, since it is twice as `wide' (i.e., the interval $[x-2, x]$ is twice as long as the interval $[x,x+1]$) and everywhere at least half as `tall' (i.e., the maximum value of $\bupdf$ on the interval $[x-2,x+1]$ is no more than twice the minimum value). 
This guarantees that by choosing $\BasicGamble$, at every point $x$ on the horizontal axis, you move more probability mass from the left of that point to the right (increasing the probability of a payoff $\geq x$) than from the right to the left (decreasing the probability of a payoff $\geq x$), which means that $\BasicGamble$ stochastically dominates $\NullOption$.\footnote{Formally, $\forall x \forall y \left( |x - y| \leq 3 \Rightarrow \frac{\bupdf(x)}{\bupdf(y)} \leq 2 \right) $ implies $\forall x \left( \int_{x-2}^{x}\bupdf(y) \, dy > \int_{x}^{x+1}\bupdf(y) \, dy \right) $, which in turn implies $\sdArgs{\BasicGamble}{\NullOption}$.}

%
%
%
%
%

 
For $\bupdf$ to never decrease by more than a factor of 2 within an interval of length 3, it is sufficient that $\bupdf$ has large tails and a high enough dispersion.
If a distribution has large tails, then for any finite 
$l$, there is \textit{some} finite $r$ such that 
$\bupdf$ never decreases by more than a factor of $r$ within an interval of length $l$. And if for $l = 3$ that factor is greater than 2, we can decrease it by `stretching' $\bupdf$ (increasingly rescaling it), so that its tails decay more slowly.


The resulting stochastic dominance relation can be 
visualized by representing each prospect with its CCDF, as in Fig.\@ \ref{figure-ccdfs}: The Basic Gamble stochastically dominates the Null Option iff its CCDF is everywhere greater.

\subsection{Sufficiency Theorem}
\label{section-mainresult}

We have now seen how background uncertainty can generate stochastic dominance.  
But how general is this phenomenon---does it depend on special and improbable conditions? In this section, we will partially answer that question by identifying a sufficient condition for $O_i$ to stochastically dominate $O_j$ under background uncertainty, 
that depends only on (i) a measure of the expectational superiority of $O_i$ to $O_j$ and (ii) the rate at which the tails of $\bupdf$ 
decay, relative to the range of possible simple payoffs from $O_i$ and $O_j$.

\begin{figure}
	\centering
%
%
%
%
	\includegraphics[width=0.3\linewidth]{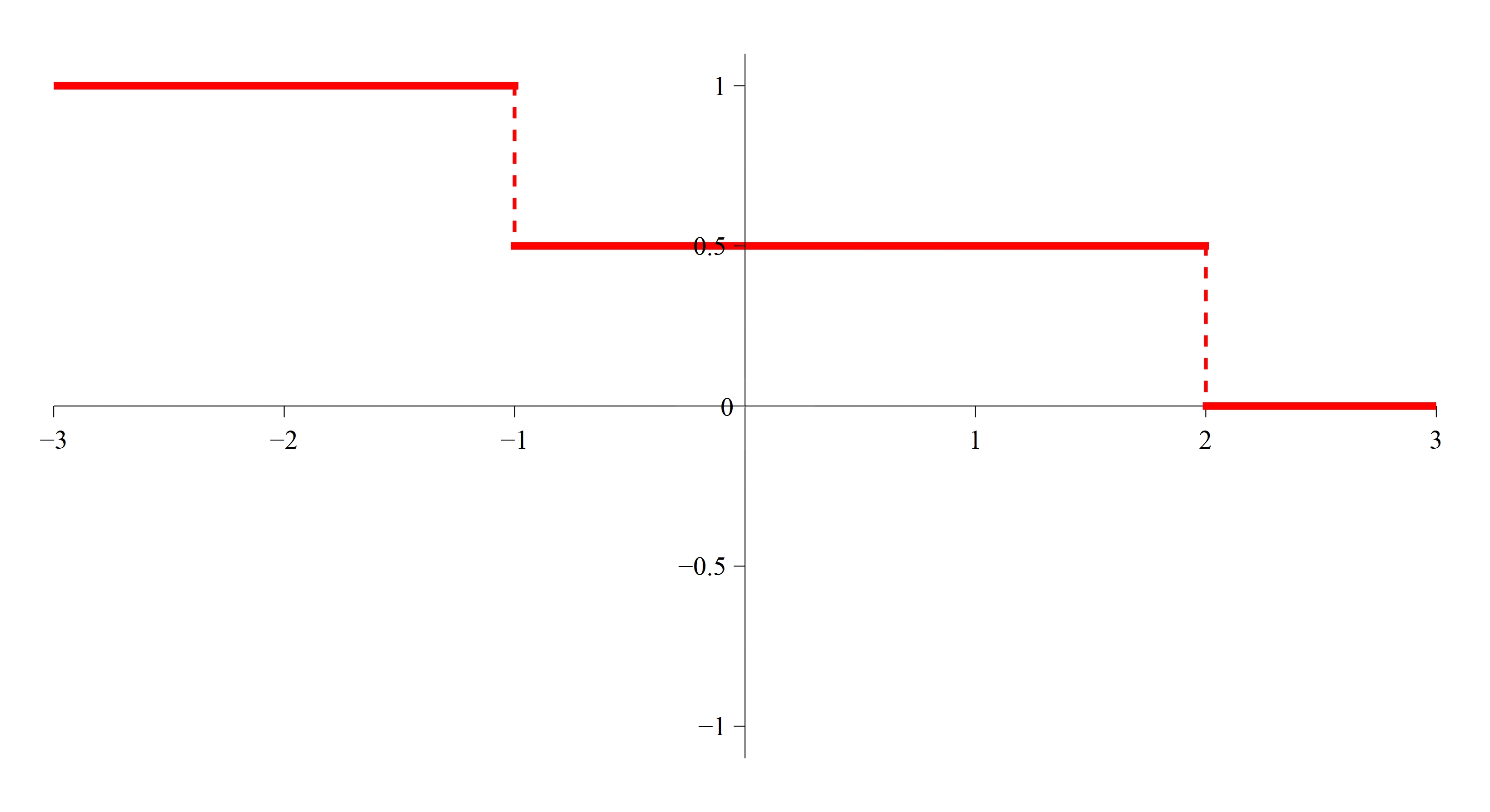}
	\includegraphics[width=0.3\linewidth]{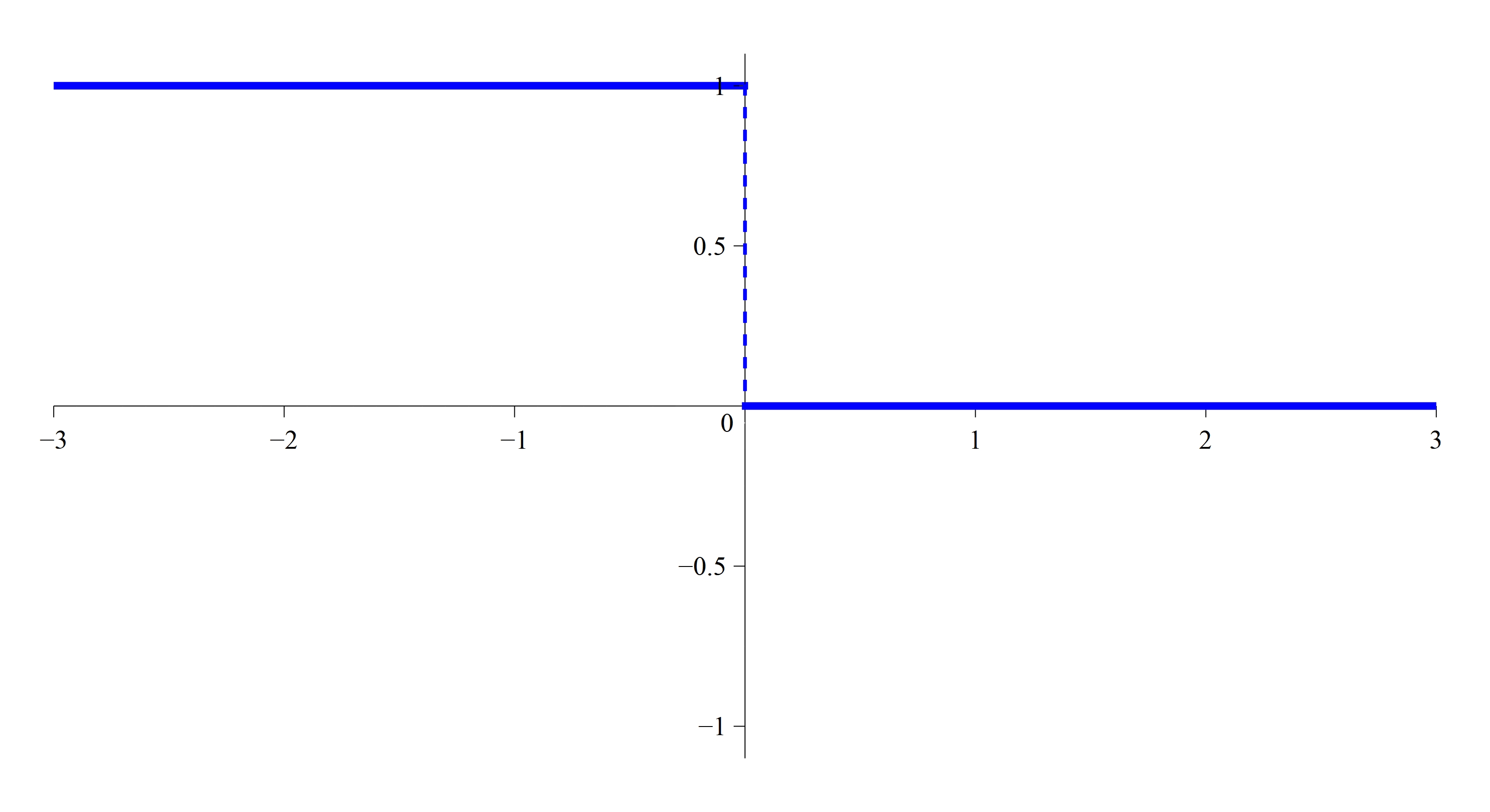}
	\includegraphics[width=0.3\linewidth]{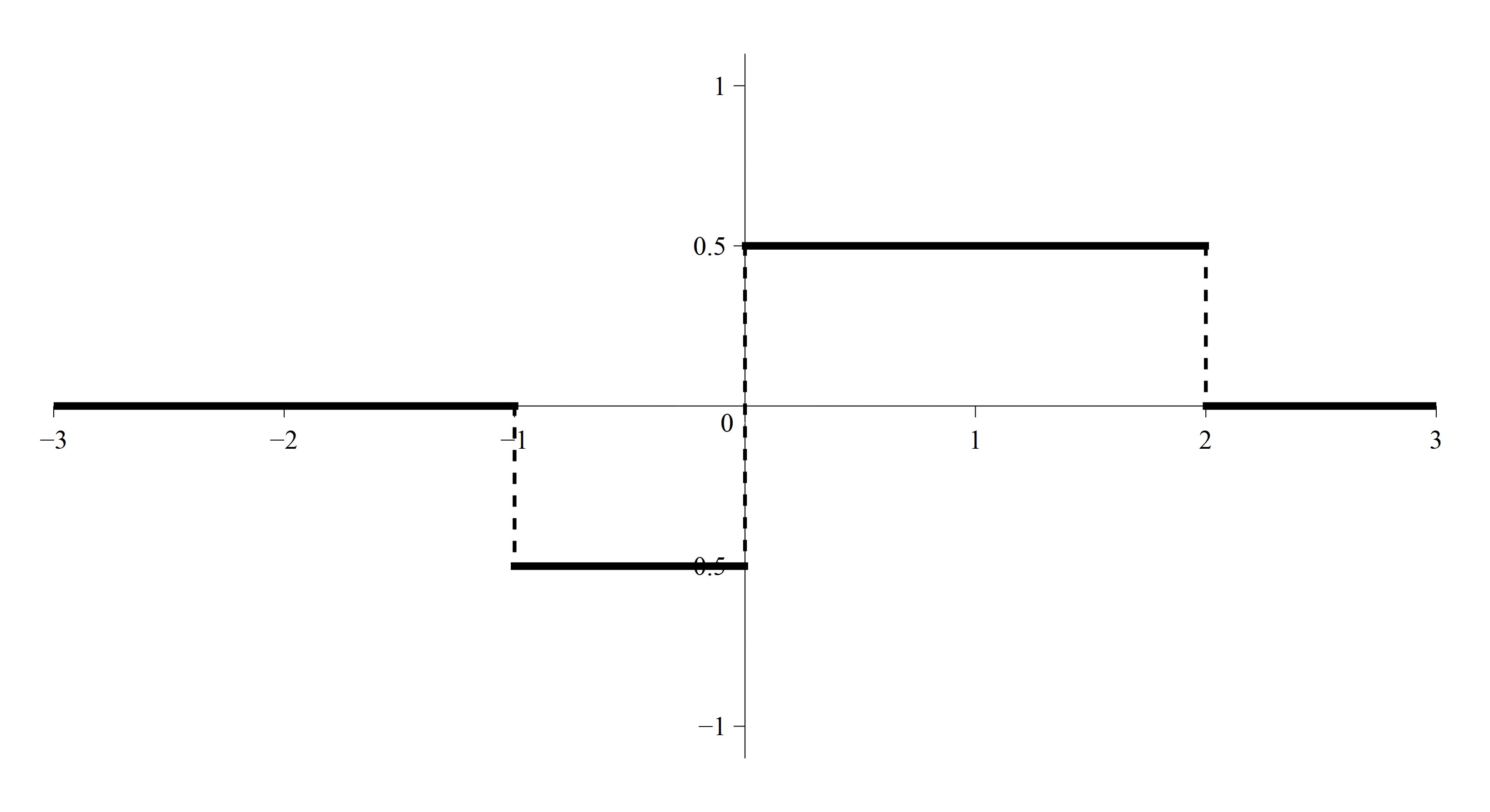}
	\caption{The CCDFs of the simple prospects of the Basic Gamble $\BasicGamble$ (left) and the Null Option $\NullOption$ (middle), and the difference $\cdfDiff{\BasicGamble}{\NullOption}$ of the two CCDFs (right).} 
	\label{figure-illustratingDelta}
\end{figure}

To state the result, we need some additional notation. 
First, we introduce a function that, for options $O_i$ and $O_j$, gives the difference between the probability that $O_i$ yields a simple payoff $\geq x$ and the probability that $O_j$ yields a simple payoff $\geq x$.

$$\cdfDiff{i}{j}(x) := \Pr(O_i \geq x) - \Pr(O_j \geq x)$$
$\cdfDiff{i}{j}$ can be understood as the difference of the CCDFs of the simple prospects of $O_i$ and $O_j$ (Fig.\@ \ref{figure-illustratingDelta}). 
We also define the positive and negative parts of $\cdfDiff{i}{j}$:

$$\cdfDiff{i}{j}^+(x) := \max(\cdfDiff{i}{j}(x),0)$$
$$\cdfDiff{i}{j}^-(x) := \max(-\cdfDiff{i}{j}(x),0)$$

The integral of $\cdfDiff{i}{j}$, $\int_{-\infty}^{\infty} \cdfDiff{i}{j}(x) \, dx = \int_{-\infty}^{\infty} \cdfDiff{i}{j}^+(x) \, dx - \int_{-\infty}^{\infty} \cdfDiff{i}{j}^-(x) \, dx$, gives the difference in expected choiceworthiness between $O_i$ and $O_j$. 
If $\cdfDiff{i}{j}$ is nowhere negative and somewhere positive, then $O_i$'s simple prospect stochastically dominates $O_j$'s, which guarantees that $O_i$ will stochastically dominate $O_j$ in any state of background uncertainty \cite[p.\@ 1880]{pomatto2020stochastic}. On the other hand, if $O_j$'s simple prospect has a greater expectation than $O_i$'s, then it is impossible for $O_i$ to stochastically dominate $O_j$ in any state of background uncertainty \cite[p.\@ 1880]{pomatto2020stochastic}. Thus, the cases of interest to us are those where $\cdfDiff{i}{j}$ is somewhere positive and somewhere negative, and where the integral of $\cdfDiff{i}{j}$ is positive. 

Second, we introduce a function $\maxRatioParam(O_i,O_j,\bupdf)$ that gives the maximum ratio between values of $\bupdf$ for arguments that differ by no more than the range of the support of $\cdfDiff{i}{j}$, denoted $|\supp(\cdfDiff{i}{j})| = \max(\supp(\cdfDiff{i}{j})) - \min(\supp(\cdfDiff{i}{j}))$. (In general, $|\supp(\cdfDiff{i}{j})|$ is the difference between the best and worst possible simple payoffs in $O_i$ and $O_j$.)

$$\maxRatioParam(O_i,O_j,\bupdf) := \max_{x, y: |y| < |\supp(\cdfDiff{i}{j})|} \frac{\bupdf(x + y)}{\bupdf(x)}$$


	
	
	
	
	
		

This notation in hand, we can now state the first result. 


\begin{restatable}[Sufficiency Theorem]{thm}{SufficiencyTheorem}
	\label{theorem-SufficiencyThm}
	
	For any options $O_i$, $O_j$ and background prospect $\bupdf$,
	
	$$\frac{\int_{-\infty}^{\infty} \cdfDiff{i}{j}^+(x) \, dx}{\int_{-\infty}^{\infty} \cdfDiff{i}{j}^-(x) \, dx} > \maxRatioParam(O_i,O_j,\bupdf) \Rightarrow \sdArgs{O_i}{O_j}.$$
\end{restatable}

%


The proof is given in the appendix. Intuitively, the theorem can be understood as follows: $\int_{-\infty}^{\infty} \cdfDiff{i}{j}^+(x) \, dx$ measures the `expectational upside' of choosing $O_i$ over $O_j$, and $\int_{-\infty}^{\infty} \cdfDiff{i}{j}^-(x) \, dx$ measures the `expectational downside', in the sense that the difference in expected choiceworthiness between $O_i$ and $O_j$ is given by $\int_{-\infty}^{\infty} \cdfDiff{i}{j}^+(x) \, dx - \int_{-\infty}^{\infty} \cdfDiff{i}{j}^-(x) \, dx$. The Sufficiency Theorem says that if the \textit{ratio} of expectational upside to expectational downside 
is greater than the maximum amount by which the value of $\bupdf$ decays over an interval of length $|\supp(\cdfDiff{i}{j})|$ (roughly, the range of possible simple payoffs from $O_i$ and $O_j$), then $O_i$ stochastically dominates $O_j$.

If $\bupdf$ has large tails, then for any $O_i$, $O_j$, $\maxRatioParam(O_i,O_j,\bupdf)$ will be finite.\footnote{Since $\frac{|\bupdf'(x)|}{\bupdf(x)}$ is bounded above by $r$, the ratio between values of $\bupdf$ separated by less than $|\supp(\cdfDiff{i}{j})|$ (i.e., $\maxRatioParam(O_i,O_j,\bupdf)$) cannot be greater than $e^{r|\supp(\cdfDiff{i}{j})|}$. This follows from the differential form of Gr\"{o}nwall's inequality \citep{gronwall1919note}.} 
Moreover, if we `stretch' $\bupdf$ along the $x$-axis (i.e., increasingly rescale it), $\maxRatioParam(O_i,O_j,\bupdf)$ converges to 1.\footnote{Rescaling $\bupdf$ by a factor of $s$ means transforming it to $\bupdf_s(x) = \frac{1}{s} \bupdf(\frac{x - a}{s})$, for some constant $a$. Comparing the corresponding points in the original and transformed distributions ($x$ and $\frac{x - a}{s}$), we find that $\bupdf(x)$ is reduced by a factor of $s$, but $\bupdf'(x)$ is reduced by a factor of $s^2$. So if $\frac{|\bupdf'(x)|}{\bupdf(x)}$ is bounded above by $r$, then $\frac{|\bupdf'_s(x)|}{\bupdf_s(x)}$ is bounded above by $\frac{r}{s}$, and $\maxRatioParam(O_i,O_j,\bupdf_s)$ is bounded above by $e^{\frac{r|\supp(\cdfDiff{i}{j})|}{s}}$. This implies that as $s$ goes to infinity, $\maxRatioParam(O_i,O_j,\bupdf_s)$ goes to $1$.} 
So if $\bupdf$ has large tails and $O_i$'s simple prospect is expectationally superior to $O_j$'s (so that $\frac{\int_{-\infty}^{\infty} \cdfDiff{i}{j}^+(x) \, dx}{\int_{-\infty}^{\infty} \cdfDiff{i}{j}^-(x) \, dx} > 1$), $O_i$ will stochastically dominate $O_j$ given a sufficient rescaling of $\bupdf$. 
This means that, as we increasingly rescale $\bupdf$ (increasing its dispersion), the partial ordering of options by stochastic dominance, $\sd$, 
asymptotically approaches the ordering of options by the expectations of their simple prospects. 

\subsection{Necessity Theorem}
\label{section-necessitythm}

The Sufficiency Theorem also offers some suggestion that it is harder for background uncertainty to generate stochastic dominance in Pascalian contexts: All else being equal, increasing the range of simple payoffs increases $\maxRatioParam(O_i,O_j,\bupdf)$, and so the condition $\frac{\int_{-\infty}^{\infty} \cdfDiff{i}{j}^+(x) \, dx}{\int_{-\infty}^{\infty} \cdfDiff{i}{j}^-(x) \, dx} > \maxRatioParam(O_i,O_j,\bupdf)$ becomes more demanding. But since this is a sufficient rather than a necessary condition for stochastic dominance, this is only a suggestion.

The suggestion is confirmed, however, by the following \textit{necessary} condition for stochastic dominance:



	
\begin{restatable}[Necessity Theorem]{thm}{NecessityTheorem}
	\label{theorem-NecessityThm}

For any options $O_i$, $O_j$ and background prospect $\bupdf$,

$$\sdArgs{O_i}{O_j} \Rightarrow \max_x \cdfDiff{i}{j}(x) > \max_x \int_{-\infty}^{\infty} \cdfDiff{i}{j}^-(x - y) \bupdf(y) \, dy.$$

\end{restatable}



%
%


The proof is again left for the appendix. Intuitively, the theorem can be understood as follows: Choosing $O_i$ rather than $O_j$ changes the probability of an overall payoff $\geq x$ by $\int_{-\infty}^{\infty} \cdfDiff{i}{j}(x - y) \bupdf(y) \, dy$, which can be decomposed into positive and negative components, as $\int_{-\infty}^{\infty} \cdfDiff{i}{j}^+(x - y) \bupdf(y) \, dy - \int_{-\infty}^{\infty} \cdfDiff{i}{j}^-(x - y) \bupdf(y) \, dy$. Call $\int_{-\infty}^{\infty} \cdfDiff{i}{j}^+(x - y) \bupdf(y) \, dy$ the \textit{increment} to the probability of a payoff $\geq x$ from choosing $O_i$ over $O_j$, and $\int_{-\infty}^{\infty} \cdfDiff{i}{j}^-(x - y) \bupdf(y) \, dy$ the \textit{decrement}. The Necessity Theorem says that, for $O_i$ to stochastically dominate $O_j$, the maximum amount by which $O_i$ increases the probability of achieving a \textit{simple} payoff $\geq x$, for any 
$x$ (given by $\max_x \cdfDiff{i}{j}(x)$), must exceed the maximum decrement to the probability of an \textit{overall} payoff $\geq x$, for any 
$x$ (given by $\max_x \int_{-\infty}^{\infty} \cdfDiff{i}{j}^-(x - y) \bupdf(y) \, dy$).


This result tells us two things. First, whenever $O_i$'s simple prospect  does not stochastically dominate $O_j$'s (so that $\max_x \int_{-\infty}^{\infty} \cdfDiff{i}{j}^-(x - y) \bupdf(y) \, dy$ is non-zero), there is some probability threshold such that sets of simple payoffs with total probability below that threshold cannot generate stochastic dominance, no matter their magnitude. To illustrate, suppose we make the choice between $O_i$ and $O_j$ more Pascalian by taking each positive simple payoff of $O_i$ and negative simple payoff of $O_j$, and replacing its simple payoff-probability pair with a smaller probability of a proportionately larger simple payoff, plus a complementary probability of a simple payoff of 0---that is, replacing $\langle v_k,p_k \rangle$ with $\langle c v_k, \frac{p_k}{c} \rangle, \langle 0, p_k - \frac{p_k}{c} \rangle$, for some constant $c$. 
This `Pascalian transformation' preserves the expectations of both options. But as $c$ goes to infinity, $\max_x \cdfDiff{i}{j}(x)$ goes to 0, while $\max_x \int_{-\infty}^{\infty} \cdfDiff{i}{j}^-(x - y) \bupdf(y) \, dy$ does not, so $O_i$ must eventually cease to stochastically dominate $O_j$. More generally, holding other features of a choice situation fixed, SDTR will eventually cease to require the expectationally superior option as the source of its expectational superiority becomes increasingly Pascalian (reliant on very small probabilities of extreme simple payoffs).\footnote{In a very specific and limited sense, therefore, SDTR vindicates the oft-mooted idea that it is permissible to ignore outcomes with sufficiently small probabilities (described, for instance, as `morally impossible' (\citeauthor{buffon1777essai}, 1777), `\textit{de minimis}' \citep{whipple2012minimis}, or `rationally negligible' \citep{smith2014evaluative}). 
But the Pascalian threshold drawn by SDTR differs from these previous ideas in important ways. First, it applies to \textit{sets} of outcomes rather than individual outcomes (so it does not face problems of individuation). 
Second, it is sensitive in precise ways to other features of the choice situation (e.g., to the magnitude of any high-probability simple payoffs that must be weighed against the more improbable outcomes, and to the dispersion of the agent's background prospect), so it does not establish any \textit{general} threshold below which probabilities can be ignored.}\ifanon\else$^{,}$\footnote{This means, among other things, that SDTR offers an appealing response to Pascal's Mugging \citep{bostrom2009pascal}. In Bostrom's case, a `mugger' asks you to hand over your wallet, promising that if you do, he will return tomorrow and use his superhuman powers to give you an enormous reward. If you are skeptical of the mugger's promise, he simply increases the extravagance of the promised reward until its expected value exceeds the value of the contents of your wallet. 
Unless your credence in the mugger's promise decreases in inverse proportion to the value of the promised reward, then as an expected value maximizer, you will eventually be 
forced to hand over your wallet. What is worrisome about this case is not merely that expectational reasoning allows your choice to be determined by minuscule probabilities of astronomical payoffs, but that it reveals an apparent vulnerability to \textit{manipulation} by other agents. An agent guided by SDTR, 
however, is significantly more resistant to this sort of manipulation. Although she is rationally \textit{permitted} to accept the mugger's offer (at least whenever it is expectation-maximizing), 
she is not rationally \textit{required} to, as long as her credence in the mugger's promise is below the `Pascalian threshold' determined by her background uncertainty and the value of her wallet. Once her credence falls below that threshold, no further increase in the extravagance of the mugger's promise can generate a rational requirement to surrender her wallet. To avoid being mouse-trapped by the mugger's ever-grander promises, the expectationalist must make the rather implausible assumption that one's credence in those promises should decrease at least inversely to the value of the promised reward. But on SDTR, even in the presence of large-tailed background uncertainty, we need make only the much weaker assumption that one's credence in the mugger's promise should decrease (at whatever rate) toward some value (not necessarily zero) below the Pascalian threshold.}\fi



But second, the probability threshold established by the Necessity Theorem---namely, $\max_x \int_{-\infty}^{\infty} \cdfDiff{i}{j}^-(x - y) \bupdf(y) \, dy$---is sensitive to the dispersion of $\bupdf$. 
As we increasingly rescale $\bupdf$ (increasing its dispersion without changing its shape), 
we spread its fixed budget of probability mass more thinly, so that $\max_x \int_{-\infty}^{\infty} \cdfDiff{i}{j}^-(x - y) \bupdf(y) \, dy$ must shrink, approaching zero in the limit.
Thus, the greater the dispersion of $\bupdf$, the more Pascalian a choice situation can become while preserving stochastic dominance. 

\subsection{Illustrations and practical implications}
\label{section-illustrationsAndImplications}

The Sufficiency and Necessity Theorems give separate sufficient and necessary conditions for stochastic dominance. If we fill in some details, though, we can find necessary-and-sufficient conditions for stochastic dominance in restricted contexts. This lets us see just how tightly SDTR constrains risk attitudes in particular choice situations, both `ordinary' and `Pascalian'.

First, let's specify a background prospect: a Laplace distribution with a mean of zero and a scale parameter of $-\frac{500}{\ln (0.05)}$ ($\approx 166.9$).\footnote{Stochastic dominance is invariant under translations of the background prospect (i.e., transformations of the form $\bupdf'(x) = \bupdf(x - a)$ for some constant $a$), since translations of $\beta$ only result in identical translations of each option's overall prospect. So 
the choice of mean 
makes no difference for our purposes.} 
A Laplace distribution has exponential tails, 
and is therefore as light in the tails as any large-tailed distribution can be. The scale parameter of $-\frac{500}{\ln (0.05)}$ is chosen because it yields a 95\% confidence interval of $[-500,+500]$, which gives an intuitive sense of the dispersion of the distribution. As we have implicitly done in previous examples, let's assume that units represent lives saved/lost---or more precisely, the choiceworthiness of saving a typical happy human life (treating this, somewhat unrealistically, as a known quantity). We can abbreviate these units as \textit{life equivalents} (LE). 
Our agent, then, is 95\% confident that her background payoff will fall in an interval whose magnitude is 1000 LE (i.e., the value of 1000 human lives). For an agent who attaches normative weight to the total value of the world that results from her choices, 
this dispersion is implausibly small (as I will argue in \S \ref{section-highDispersion}). 
But I choose it 
in order to emphasize how easily background uncertainty can generate very strong stochastic dominance constraints on an agent's choices.


To see the strength of these constraints, consider the following: 

\begin{description}
	\item[ ] \textbf{Generalized Basic Gamble (\textit{G}$\mathbf{'}$)} $\{ \langle -1 , 0.5 \rangle, \langle 0 , 0.5 - p \rangle, \langle 2 , p \in (0,0.5] \rangle \}$
\end{description}
We can interpret $G'$ as an option that will save two lives with probability $p$, cause one death with probability $0.5$, and have no consequences with probability $0.5 - p$.

$G'$ has greater expected choiceworthiness than the Null Option $\NullOption$, of course, iff $p > 0.25$. By comparison, and somewhat surprisingly, 
$G'$ stochastically dominates $\NullOption$ iff $p >$ $\sim 0.25226$.\footnote{Consider the CDF of the background prospect: 
	
	$$\bucdf(x) = \begin{cases} 
	0.5\exp(\frac{\ln(0.05)x}{500}) & x\leq 0 \\
	1 - 0.5\exp(-\frac{\ln(0.05)x}{500}) & x > 0
	\end{cases}$$
	
For any \textit{x}, $G'$ improves the probability of a payoff $\geq x$ (relative to $\NullOption$) by $p(\bucdf(x) - \bucdf(x-2))$, and worsens the probability of a payoff $\geq x$ (relative to $\NullOption$) by $0.5(\bucdf(x+1) - \bucdf(x))$. Thus, $G'$ stochastically dominates $\NullOption$ iff $\forall x (p(\bucdf(x) - \bucdf(x-2)) > 0.5(\bucdf(x+1) - \bucdf(x)))$, or equivalently,  $\forall x \left( p > \frac{ 0.5(\bucdf(x+1) - \bucdf(x)) }{\bucdf(x) - \bucdf(x-2)} \right) $. And the function on the right-hand side of this inequality is bounded above at 
$\sim 0.25226$. 
} So, even given a relatively light-tailed background prospect with modest dispersion, stochastic dominance imposes \textit{extremely} tight constraints on the choice between $G'$ and $\NullOption$---nearly as tight as those imposed by expectationalism.

This example illustrates the following more general point. If $\bupdf$ has large tails, then under various reasonable assumptions about its shape (e.g., that it belongs to a standard parameterized family of distributions like Laplace or Cauchy), 
SDTR will be tightly constraining (closely approximating the ranking of options by the expectations of their simple prospects) in any choice situation where the interquartile range 
of $\bupdf$ is significantly greater than the range of possible simple payoffs. 
Specifically, in these circumstances, $\maxRatioParam(O_i,O_j,\bupdf)$ will be not much greater than 1, and thus the sufficient condition for stochastic dominance given by the Sufficiency Theorem will be relatively easily met. For instance, if $\bupdf$ is a Laplace distribution and its $\IQR$ is ten times greater than $|\supp(\cdfDiff{i}{j})|$ (which, remember, is less than or equal to the range of possible simple payoffs from $O_i$ and $O_j$), then $\maxRatioParam(O_i,O_j,\bupdf) \approx 1.15$. For a Cauchy distribution, the equivalent figure is $\sim 1.22$. 
And as per the Sufficiency Theorem, it is sufficient (though not necessary) for stochastic dominance that $\frac{\int_{-\infty}^{\infty} \cdfDiff{i}{j}^+(x) \, dx}{\int_{-\infty}^{\infty} \cdfDiff{i}{j}^-(x) \, dx}$ (the ratio of `expectational upside' to `expectational downside' from choosing $O_i$ over $O_j$) exceeds this threshold.


By contrast, consider the following 
`Pascalian transformation' of $G'$:

\begin{description}
	\item[ ] \textbf{Generalized Pascalian Gamble (\textit{G}$\mathbf{''}$)} $\{ \langle -1 , 0.5 \rangle, \langle 0 , 0.5 - p \rangle, \langle 2000 , p \in (0,0.0005] \rangle \}$
\end{description}

$G''$ has greater expected choiceworthiness than $\NullOption$ iff $p > 0.00025$. But in this case, $G''$ only comes to stochastically dominate $\NullOption$ when $p >$ $\sim 0.0030047$---\textit{more than ten times} the probability at which $G''$ becomes expectationally superior.\footnote{By reasoning parallel to the case of $G'$, $G''$ stochastically dominates $\NullOption$ iff $\forall x \left( p > \frac{ 0.5(\bucdf(x+1) - \bucdf(x)) }{\bucdf(x) - \bucdf(x-2000)} \right) $. And the function on the right-hand side of this inequality is bounded above at $\sim 0.0030047$.} 
This illustrates the difference between the tight constraints imposed by stochastic dominance in cases involving intermediate probabilities of modest simple payoffs, and the relative latitude it allows 
in cases involving very small probabilities of very large simple payoffs.
\footnote{Notably, given that $\bupdf$ has large tails, it seems to matter very little precisely how heavy its tails are. For instance, suppose we replace the Laplace distribution with a Cauchy distribution (which has \textit{much} heavier tails) 
with a scale parameter of $-500(\cot(0.525\pi))$ ($\approx 39.35$)---which yields the same 95\% confidence interval of $[-500,+500]$. Now we find that $G'$ stochastically dominates $\NullOption$ iff $p > \sim 0.25969$ (as opposed to $\sim 0.25226$ for the Laplace distribution), and $G''$ stochastically dominates $\NullOption$ iff $p > \sim 0.009452$  (as opposed to $\sim 0.0030047$ for the Laplace distribution). 
So at least in these two cases, moving to a much heavier-tailed background prospect 
with 
similar dispersion does not change the conditions for stochastic dominance very much, and in fact makes those conditions somewhat \textit{more} demanding.} 


What does this mean for potentially Pascalian choices in the real world---e.g., choosing between interventions that do moderate amounts of good with high probability in the near term and interventions that try to influence the far future, doing potentially astronomical good, but with (plausibly) very low probability of success?
Fully answering this question is a large project unto itself (requiring, among other things, a plausible model of our actual background uncertainty and of the probabilities and payoffs involved in the interventions we wish to compare). But as a first approximation, let's consider another stylized case in which we must choose between a `sure thing' option that saves some small number of lives $s$ for certain ($S = \{ \langle s, 1\rangle \})$ and an expectationally superior `long shot' option that tries to prevent existential catastrophe, thereby enabling the existence of astronomically many future lives, but has only a very small probability of making any difference at all ($L = \{\langle 0, 1 - p \rangle, \langle a, p \rangle\}$, where $a$ is astronomically large, $p$ is very small, and $ap > s$). And let's assume of the agent's background prospect only that it has large tails and that its dispersion (as measured by interquartile range) is several orders of magnitude greater than the sure-thing payoff $s$.

First, the Necessity Theorem implies that there is a minimum value of $p$ below which $L$ cannot stochastically dominate $S$, no matter the magnitude of $a$. 
It turns out that this threshold can be fairly well approximated by $\frac{s}{\IQR(\bupdf)}$, the ratio of the sure-thing payoff to the interquartile range of the background prospect.\footnote{Recall that by the Necessity Theorem, $L \sd S$ under background prospect $\bupdf$ only if $\max_x \cdfDiff{L}{S}(x) > \max_x \int_{-\infty}^{\infty} \cdfDiff{L}{S}^-(x - y) \bupdf(y) \, dy$. In this case, $\max_x \cdfDiff{L}{S}(x) = p$, so this is equivalent to $p > \max_x \int_{-\infty}^{\infty} \cdfDiff{L}{S}^-(x - y) \bupdf(y) \, dy$. $\cdfDiff{L}{S}^-$ is simply a `rectangular' function with $\cdfDiff{L}{S}^-(x) = 1-p$ on the interval $(0,s]$ and 0 elsewhere. So $\max_x \int_{-\infty}^{\infty} \cdfDiff{L}{S}^-(x - y) \bupdf(y) \, dy = \int_{a}^{a + s} (1-p) \bupdf(y) \, dy$, for $a$ such that $[a, a+s]$ is the interval of length $s$ to which $\bupdf$ assigns greatest probability.
	
Since $\IQR(\bupdf) > s$, the average value 
of $\bupdf$ on the interval $[a, a + s]$ must be at least as great as its average value 
on its interquartile interval, $\frac{0.5}{\IQR(\bupdf)}$. So $\int_{a}^{a + s} (1-p) \bupdf(y) \, dy \geq s (1-p) \frac{0.5}{\IQR(\bupdf)}$. On the other hand, the average value 
of $\bupdf$ on any interval cannot exceed its maximum value, $\max_x\bupdf(x)$. So $\int_{a}^{a + s} (1-p) \bupdf(y) \, dy \leq s (1-p) \max_x\bupdf(x)$. Combining these observations, we conclude that $\int_{a}^{a + s} (1-p) \bupdf(y) \, dy$  (and therefore $\max_x \int_{-\infty}^{\infty} \cdfDiff{L}{S}^-(x - y) \bupdf(y) \, dy$) must be in the interval $[s (1-p) \frac{0.5}{\IQR(\bupdf)}, s (1-p) \max_x\bupdf(x)]$.
	
With a little rearrangement, this implies that the `Pascalian threshold' given by the Necessity Theorem (i.e., the minimum value of $p$ below which $L$ cannot stochastically dominate $S$ under background prospect $\bupdf$) 
is in the interval $[\frac{0.5s}{\IQR(\bupdf)} ( 1 + \frac{0.5s}{\IQR(\bupdf)} )^{-1}, s \max_x\bupdf(x) (1 + s \max_x\bupdf(x))^{-1}]$. Typically, $\max_x\bupdf(x)$ will be not much greater than $\frac{0.5}{\IQR(\bupdf)}$ (the average value of $\bupdf$ over its interquartile interval). (For Laplace distributions, for instance, the ratio of $\max_x\bupdf(x)$ to $\frac{0.5}{\IQR(\bupdf)}$ is $2\ln 2 \approx 1.39$; for Cauchy distributions, it is $\frac{4}{\pi} \approx 1.27$.) Combining this observation with the stipulation that $s \ll \IQR(\bupdf)$, we can conclude that both $( 1 + \frac{0.5s}{\IQR(\bupdf)} )^{-1}$ and $(1 + s \max_x\bupdf(x))^{-1}$ are very close to 1, and therefore that the Pascalian threshold is either within or at most very slightly below the interval  $[\frac{0.5s}{\IQR(\bupdf)}, s \max_x\bupdf(x)]$. And this means, for instance, that as long as $\max_x\bupdf(x)$ does not exceed $\frac{0.5}{\IQR(\bupdf)}$ by more than a factor of 
4, the Pascalian threshold will be approximated by $\frac{s}{\IQR(\bupdf)}$ to within roughly a factor of $2$.}
So for instance, if $s = 10$ LE and $\IQR(\bupdf) = 10^9$ LE, 
then the `Pascalian threshold' for values of $p$ below which $L$ cannot stochastically dominate $S$ will likely be in the neighborhood of $10^{-8}$ (with its exact value depending on the shape of $\bupdf$).

This threshold applies to $L$ \textit{no matter the magnitude} of the astronomical simple payoff $a$. How much do things change if we consider some particular value of $a$, like Bostrom's $10^{52}$ LE? The short answer is: not much. For these purposes, payoffs that are very large relative to the dispersion of the background prospect can, to a very close approximation, be treated as infinite.\footnote{To see this, consider $L_{1}: \{ \langle 0, 1 - p \rangle, \langle 10^{52}, p \rangle \}$ and $L_{2}: \{ \langle 0, 1 - p \rangle, \langle + \infty, p \rangle \}$. Will $L_{2}$ stochastically dominate $S$ for much smaller values of $p$ than $L_{1}$? $L_{1} \sd S$ iff $\forall x p \int_{x - 10^{52}}^{x} \bupdf(y) \, dy \geq \int_{x - s}^{x} \bupdf(y) \, dy$. As long as $p > 10^{-52}s$, this condition will be satisfied for values of $x$ in the right tail of the background prospect. (In particular, if $\bupdf$ is unimodal and symmetrical, then clearly $\forall x p \int_{x - 10^{52}}^{x} \bupdf(y) \, dy \geq \int_{x - s}^{x} \bupdf(y) \, dy$ for values of $x$ that exceed the mode of $\bupdf$ by at least $0.5 \times 10^{52}$.) 
For all other values of $x$, $\int_{x - 10^{52}}^{x} \bupdf(y) \, dy$ is only \textit{very} slightly smaller than $\int_{- \infty}^{x} \bupdf(y) \, dy$, assuming 
$\IQR(\bupdf) \ll 10^{52}$, since the lower integration bound $x - 10^{52}$ will be far out in the left tail of $\bupdf$. Thus, the value of $p$ required for $L_{1}$ to stochastically dominate $S$ is only very slightly greater than the value required by $L_{2}$.} Thus, if $a = 10^{52}$ LE and $\IQR(\bupdf) \ll 10^{52}$ LE, the threshold at which $L$ stochastically dominates $S$ will be roughly $\frac{s}{\IQR(\bupdf)}$.


This suggests a strategy for proponents of Bostrom-style arguments (and `longtermists' more generally) to allay concerns about Pascalian fanaticism. Suppose you have some limited resource, like money, that you can use either to do some definite short-term good or to slightly increase the probability of a positive long-term trajectory for humanity. 
And suppose we know the latter option to be expectationally superior, despite its small probability of impact. If the amount by which a marginal unit of resource can increase the probability of a positive long-term trajectory exceeds the ratio between the amount of good you could do by spending that same unit of resource on short-term causes and the interquartile range of your background prospect, then very plausibly the longtermist option will turn out to be stochastically dominant, 
in which case we should have no decision-theoretic reservations about favoring it. 
For instance, suppose you can either save 100 lives for sure or reduce the probability of existential catastrophe by $10^{-9}$, thereby potentially 
enabling $10^{52}$ future lives. If the 
interquartile range of your background prospect is significantly greater than $10^{11}$ LE, then the longtermist option is likely to be stochastically dominant. And if your background uncertainty reflects uncertainty about the total amount of value in the Universe, it seems quite plausible---though perhaps not indisputable---that its dispersion should be at least this great. 
(We will consider this question in \S \ref{section-highDispersion}.) But if, on the other hand, you can only have a much smaller effect on the probability of existential catastrophe (say, $10^{-30}$), then much greater background uncertainty will be needed for stochastic dominance, and even though the expectations may still be astronomical (in this case, $10^{22}$ LE), it looks 
more plausible that you are rationally permitted to prefer the expectationally inferior `sure thing'.\footnote{For a general exposition of the case for longtermism (roughly, the thesis that what we ought to do is primarily determined by the effects of our choices on the far future) based on the potentially astronomical scale of future 
civilization, see \cite{beckstead2013overwhelming,beckstead2019brief} and \cite{greavesMScase}. For discussion of the worry that these `astronomical stakes' arguments involve a problematic form of Pascalian fanaticism, see Chs.\@ 6--7 of \cite{beckstead2013overwhelming}.} 


In light of the preceding discussion, it seems to me that the greatest intuitive worry about SDTR in the presence of large-tailed background uncertainty is not that it will capture too little of expectational reasoning (failing to recover intuitive constraints on our choices), but rather that it will capture too much---requiring us to accept many gambles that seem intuitively Pascalian (e.g., where the probability of any positive payoff is on the order of $10^{-9}$ or less). 
But really, this is not a worry at all: Unlike 
primitive expectationalism, SDR is supported by \textit{a priori} arguments far more epistemically powerful than our intuitions about Pascalian gambles. If some gambles that seem intuitively Pascalian turn out to be stochastically dominant 
once we account for our background uncertainty, we should not conclude that stochastic dominance is implausibly strong. Rather, we should conclude that there is a \textit{much more compelling argument} for choosing the expectation-maximizing option in these cases than we had previously realized. This would be not a \textit{reductio} but rather an unexpected and practically important discovery.

\section{Sources of background uncertainty}
\label{section-sourcesofbackgrounduncertainty}



The 
results above  are practically significant only if some agents are (or 
ought to be) in a state of background uncertainty with large tails 
and at least moderate dispersion. In this section, I argue that this sort of background uncertainty is rationally required, at least for many agents, in real-world choice situations. In \S\S \ref{section-largeTails}--\ref{section-highDispersion}, I argue that large tails and high dispersion respectively are appropriate for aggregative consequentialists. In \S \ref{section-mixedTheories}, I generalize these arguments to agents who accept `mixed' 
theories, giving weight to overall consequences 
but also to other kinds of normative considerations. In \S \ref{section-parochialTheories}, I argue that the preceding conclusions at least partly 
generalize to `parochial' agents who give no weight at all to the value of the world as a whole, caring only about 
some narrow circle of moral concern like their family, village, or nation.


\subsection{Large tails}
\label{section-largeTails}

In this subsection, I give three arguments that our uncertainty about the total amount of value in the world has large tails, and therefore that aggregative consequentialists at least should be in a state of large-tailed background uncertainty.

First, an intuitive argument: The `large tails' condition 
is in fact very modest. Setting aside some contrived exceptions, a distribution has 
large tails as long as there is some finite upper bound on the ratios of 
probabilities assigned to adjacent intervals of a fixed length, like $[x-1,x]$ and $[x,x+1]$. 
In our context, this means there is an upper bound on \textit{how much more probable I take it to be} that the Universe contains between $x-1$ and $x$ units of value than that it contains between $x$ and $x + 1$ units of value (or vice versa). 
The only way there could fail to be such a bound (given that $\bupdf$ is supported everywhere) is if the ratio of probabilities assigned to adjacent intervals increased without bound in one or both tails of $\bupdf$. But this implies that I become arbitrarily confident about the relative probability of \textit{very} similar hypotheses, in a domain where I seem to have virtually no grounds for such discrimination. 
It would mean, for instance, that I find it \textit{vastly} more probable that the Universe contains between $-18,946,867,974,834$ and $-18,946,867,974,835$ units of value than that it contains between $-18,946,867,974,835$ and $-18,946,867,974,836$ units of value. And as the numbers get larger, my relative confidence only gets (boundlessly) greater. 
But it seems obvious that, if anything, the degree to which I discriminate between such adjacent hypotheses 
should \textit{diminish} in the extreme tails of $\bupdf$. 
None of my evidence provides any serious support for the first of the above hypotheses ($[-...5,-...4]$) over the second ($[-...6,-...5]$), at least not in any way that I am capable of identifying. 

The second argument is 
more concrete: Attempting to model our actual background uncertainty, even on fairly conservative assumptions, yields tails significantly heavier than exponential.

Assume that welfare is one of the things that contributes to the total 
amount of value in the Universe.
Then, unless other normative considerations are systematically anti-correlated with 
welfare, our background uncertainty 
should be at least as great as our uncertainty about total welfare. 
This uncertainty derives from uncertainty about both (i) the 
number of welfare subjects in the Universe and (ii) their average welfare. 
Either of these factors could give us large tails, 
but (i) is the most straightforward.

Our uncertainty about the \textit{size} of the Universe provides a useful 
lower bound on our uncertainty about the total number of welfare subjects.\footnote{It is only a lower bound because we are also very uncertain about the number of welfare subjects per unit of comoving spatial volume---see for instance \cite{sandberg2018dissolving}.} There is no known upper bound on the size of the Universe as a whole, which we know must be many times larger than the \textit{observable} universe.\footnote{Assuming that the Universe has the simplest (viz., simply connected) topology, it is finite if and only if it has positive curvature, with larger curvature implying a smaller Universe. Current cosmological data constrain the curvature of the Universe to a fairly small interval around zero \citep{gong2011current,jimenez2018peering}. Based on these data, \cite{vardanyan2011applications} find a lower bound on the size of the Universe of 251 Hubble volumes (roughly 7.7 times larger than the observable universe), with 99\% confidence. Much larger numbers have been suggested as well: \cite{greene2004fabric} notes that in many inflationary models, the Universe is so large that `[i]f the entire cosmos were scaled down to the size of earth, the part accessible to us would be much smaller than a grain of sand' (p.\@ 285). From one such inflationary model, \cite{page2007susskinds} extrapolates (though without fully endorsing) a lower bound of roughly $10^{10^{10^{122}}}$ Hubble volumes.

To my knowledge, no cosmologist has proposed an \textit{upper} bound on the size of the Universe. \cite{vardanyan2009flat} give a probability distribution that is bounded above at roughly $10^8$ Hubble volumes (p.\@ 438). But this is an artifact of their choice of categories: Because a universe larger than that bound is observationally indistinguishable from a flat (infinite) universe, they group larger finite universes together with infinite universes for purposes of model comparison (see \S 3.3, pp.\@ 435-6). \ifanon\else

I am setting aside, as overkill, various multiverse hypotheses according to which the result of the Big Bang (our observable universe, and what lies beyond it) is only a small part of the Universe as a whole. But these hypotheses of course add to our uncertainty about the size of the Universe and the total amount of value it contains.\fi} 
There is therefore no known upper bound on the number of welfare subjects in the Universe. Actually quantifying our uncertainty about the size of the Universe requires a choice of prior, which is of course  a fraught endeavor whose philosophical difficulties we will not be able to resolve here. But the best we can do is to choose a reasonable and conservative prior and see where it leads us. \cite{vardanyan2009flat} suggest a physically motivated prior that they call the \textit{astronomer's prior}. Conditional on a finite universe, the astronomer's prior is uniform over values of $\Omega_k$ in the interval $(0,1]$, where $\Omega_k$ is the curvature parameter in the standard $\Lambda$CDM cosmology (smaller values of $\Omega_k$ indicating less curvature and hence a larger Universe).\footnote{For motivation of the astronomer's prior, see \citeauthor{vardanyan2009flat} (\citeyear{vardanyan2009flat}, p.\@ 436). Vardanyan et al also consider a second prior, which is log-uniform over $\Omega_k$. But the plausibility of this prior depends significantly on their decision to group models with $|\Omega_k| \leq 10^{-5}$ together with $\Omega_k = 0$, since a log-uniform prior on the full interval $(0,1]$ would be improper. If we were willing to entertain this improper prior, it would yield an even heavier-tailed distribution with respect to the size of the Universe than the astronomer's prior. 
} This implies 
a prior over the present curvature radius of the Universe, $a_0$, where $\Pr(a_0) \propto a_0^{-3}$,  
which in turn implies a prior over the present \textit{volume} of the Universe, $V$, where $\Pr(V) \propto V^{-\frac{5}{3}}$.\footnote{Specifically, the astronomer's prior corresponds to the following prior over $V$:
	
$$f_V(x) = \begin{cases} 
	\frac{2^{\frac{5}{3}}}{3} \pi^{\frac{4}{3}} c^2 H_0^{-2} x^{-\frac{5}{3}} & x \geq 2 \pi^2 H_0^{-3}c^3 \\
	0 & \mathrm{otherwise}
	\end{cases}
$$
where $H_0$ is the Hubble constant and $c$ is the speed of light.
	
Of course, this is only a prior, and what we are really care about is the posterior, i.e., the probability distribution we should actually adopt given our current evidence. But since observational evidence cannot measure $\Omega_k$ to a precision greater than $\sim 10^{-4}$, it cannot discriminate within the tail of very large finite universes (corresponding to values of $\Omega_k$ asymptotically approaching zero from below), and hence cannot significantly change the tail properties of the distribution.} And this distribution is extremely heavy-tailed---much heavier than exponential. 



Given such a heavy-tailed distribution for 
the size of the Universe, a large-tailed distribution for total welfare in the Universe (or, in the part of the Universe unaffected by our choices) is nearly a foregone conclusion, requiring only (i) that the number of welfare subjects per unit of comoving spatial volume is not strongly anti-correlated with the size of the Universe\footnote{`Comoving' spatial volume is measured using present-day distances, so that the total comoving volume of the Universe remains constant over time despite cosmic expansion. `Welfare subjects per unit comoving spatial volume' can be thought of, without much loss of accuracy, as `welfare subjects per galaxy', since the comoving density of galaxies 
is more or less a known quantity.} and (ii) that we assign non-zero probability to both positive and negative values for the average welfare of all welfare subjects. 
The second assumption looks unassailable, and I cannot think of a 
reason to question the first. 

The most serious objection I can see to the preceding line of argument is that the Universe, and the number of welfare subjects it contains, may well be \textit{infinite} 
\citep{knobe2006philosophical,vardanyan2009flat,carroll2017why}. I will, unfortunately, have little to say about this issue (though I say a bit in \S \ref{section-infiniteworlds} below). I take it for granted that the true axiology 
can make non-trivial comparisons between infinite worlds, 
so that even if we were certain that the Universe was infinite, we could still be uncertain about its overall value. 
But how (if at all) we extend the arguments of this paper to the infinite context depends very much on what sort of infinite axiology we adopt, 
and there is as yet no agreement even on very basic questions about how to formulate an infinite axiology.\footnote{For some of the many extant proposals, see for instance \cite{vallentyne1997infinite}, \cite{mulgan2002transcending}, \cite{bostrom2011infinite}, and \cite{arntzenius2014utilitarianism}.\label{footnote-infiniteethicscites}} 
Perhaps more to the point (though no more satisfying), expectational reasoning is if anything \textit{more} threatened by infinite worlds than stochastic dominance reasoning (see for instance \citeauthor{bostrom2011infinite} (\citeyear{bostrom2011infinite}, pp.\@ 13ff), \cite{arntzenius2014utilitarianism}). So even if the arguments in this paper suffer in an infinitary context, that is not likely to generate much support for expectationalism over SDTR. 

The third and final argument for large tails is the simplest: When I am uncertain which of several probability distributions best characterizes some phenomenon, the resulting mixture distribution (the probability-weighted average of the distributions over which I'm uncertain) inherits the tail properties of the heaviest-tailed distribution in the mixture. (The further out we go in the tails of the mixture distribution, the more the heaviest-tailed distributions dominate the mixture.) 
So, suppose I am unsure what background prospect is justified by my evidence, or that I assign credence to multiple physical theories 
that imply different objective probability distributions over background payoffs. As long as I assign positive credence to any distribution with exponential or heavier tails, the resulting background prospect will 
have exponential or heavier tails. And, excluding some contrived and unlikely cases, a background prospect with exponential or heavier tails (i.e., that is bounded below in the tails by a Laplace distribution) 
will also satisfy our `large tails' condition. 





\subsection{High dispersion}
\label{section-highDispersion}

Large tails create the conditions for background uncertainty to generate stochastic dominance, but as we saw in \S \ref{section-results}, how closely stochastic dominance approximates the ordering of options by the expectations of their simple prospects depends on the \textit{dispersion} of the background prospect---intuitively, how `spread out' it is. A distribution can have very heavy tails while nevertheless having arbitrarily low dispersion, concentrating most of its probability mass in a very small interval. 

As we saw in \S \ref{section-illustrationsAndImplications}, SDTR is generally tightly constraining when 
$\bupdf$ has large tails and a dispersion that is large relative to the range of possible simple payoffs. 
For SDTR to yield 
intuitively satisfactory 
constraints 
in real-world choice situations, then, it should be the case that the dispersion of our real-world background uncertainty is large relative to the stakes we face in most real-world choice situations. 
If our background uncertainty reflects our uncertainty about the total amount of value in the Universe, this `high dispersion' premise strikes me as nearly indisputable, and less in need of defense than the `large tails' premise. So I will just briefly note three arguments for high dispersion. 

First, the population of welfare subjects on Earth up to the present is both very large and a matter of great uncertainty.\footnote{The historical human population is estimated at roughly $10^{11}$ \citep{kaneda2011how}. The present mammal population seems to be at least $10^{11}$, and the present vertebrate population at least $10^{13}$ \citep{tomasik2019how}, which suggests historical populations of more than $10^{18}$ and $10^{21}$ respectively. And our uncertainties even about \textit{present} mammal and vertebrate population sizes span multiple orders of magnitude \citep{tomasik2019how}. The possibility that some invertebrates (e.g., insects) are welfare subjects as well only adds to our uncertainty.} And we can say very little about average welfare in this historical population, including whether it has been positive or negative. 
Hence our uncertainty about total welfare on Earth up to the present moment (one contributor to our background uncertainty) has very high dispersion.

Second, we saw above that our uncertainty about the size of the Universe as a whole is very great \citep{vardanyan2009flat,vardanyan2011applications}, as is our uncertainty about the number of inhabited planets and hence the number of welfare subjects per unit of spatial volume \citep{sandberg2018dissolving}.\ifanon \space \else\footnote{Indeed, our uncertainty about the number of welfare subjects is much greater even than our uncertainty about the number of inhabited planets, since we know very little about the number of welfare subjects per inhabited planet---particularly in the case of 
planets that give rise to advanced civilizations.} \fi
And here too, even more so than in the terrestrial case, we know next to nothing about average welfare. So our uncertainty about the total welfare of non-Earth-originating welfare subjects outside our future light cone (another contributor to our background uncertainty) also has very high dispersion.

Finally, it is worth mentioning a third source of background uncertainty: 
an agent's uncertainty about the outcomes of other future choices (and perhaps some past choices as well), both her own and those of other agents. When those choices, and their outcomes, are suitably independent of the agent's present choice, they can be treated as part of her background uncertainty. Simplifying considerably, if an agent believes that future agents (herself perhaps included) will face $n$ choices roughly similar to her own present choice, and that these choices and their outcomes are all suitably independent, then her background uncertainty about the cumulative outcome of all those future choices will be her uncertainty about the outcome of an $n$-step independent random walk, with dispersion proportionate to $\sqrt{n}$. 
Assuming, then, that the choice our agent faces is `ordinary' enough that she expects many similar choices to be faced by 
future agents, her uncertainty about the cumulative outcome of these future choices (a third contributor to her background uncertainty) is likely to have a dispersion that is large relative to the 
possible simple payoffs of her present options.\footnote{It is worth noting that such long runs of future choices cannot on their own generate \textit{large tails}, if the simple prospects of the options in those future choices all have finite support. But they can serve to increase the dispersion of an already large-tailed background prospect.} 


To stake out a 
definite, quantitative conclusion: 
Based on the preceding arguments, it seems clear to me that the dispersion (as measured by $\IQR$) of our background uncertainty about the total amount of value in the Universe 
must be \textit{at least} on the order of $10^9$ LE, and perhaps very much larger. 

\subsection{`Mixed' normative theories}
\label{section-mixedTheories}

We have thus far focused on 
agents who are exclusively concerned with aggregative consequentialist considerations, i.e., who measure the choiceworthiness of their options entirely 
by the total amount of value in the resulting world. But I think that the results in \S \ref{section-results} are much more widely applicable. Unfortunately, I don't see any way of establishing this except 
to consider, case by case, the various kinds of normative theory 
that an agent might use to gauge the choiceworthiness of her options. And 
doing this at all comprehensively would be a long and tedious exercise. So for now, I will simply point out two ways in which the phenomenon of large-tailed background uncertainty with high dispersion, and hence the practical 
significance of the results in \S \ref{section-results}, generalizes beyond aggregative consequentialism.

First, consider an agent who accepts a `mixed' theory, measuring the choiceworthiness of her options partly by the total amount of value in the resulting world, but also by various other considerations, like whether she is keeping her promises, being a good friend, 
or exhibiting virtues like temperance and modesty, and perhaps giving extra 
weight to her own welfare/life projects, beyond their contribution to the overall value of the world. 
For simplicity, suppose 
she treats all these considerations as making additively separable contributions to the choiceworthiness of her options. And suppose that all of the non-aggregative-consequentialist considerations are captured by the simple payoffs of her options, 
so that her background uncertainty still reflects only 
uncertainty about the total amount of value in the Universe. (These are significant assumptions, though I cannot see any particular way in which plausibly weakening them would derail the conclusions below.) Then, of course, all the arguments for large tails from \S \ref{section-largeTails} will still apply. And the dispersion of $\bupdf$ 
will still typically be large relative to the range of simple payoffs, so long as it is large relative to the weight the agent attaches to ordinary non-consequentialist considerations. Suppose, for instance, that $\IQR(\bupdf) = 10^9$ LE, which I suggested above as a conservative lower bound on our uncertainty about the total amount of value in the Universe. 
Only the most extreme deontological views assign ordinary non-consequentialist considerations a weight anywhere near that order of magnitude. So this agent will still, at least in 
most choice situations, find that the dispersion of her background prospect is much larger than the possible simple payoffs of her options, and hence 
that her choices are fairly tightly constrained by stochastic dominance.


\subsection{`Parochial' normative theories}
\label{section-parochialTheories}

Next, consider an agent whose concerns are `parochial'---i.e., 
not with the value of the Universe as a whole, but 
circumscribed within some fairly narrow `moral circle'. At the most extreme, such an agent might assign normative significance only to her own welfare. Alternatively, she might assign 
significance only to the welfare of her family, her tribe, local community, or nation, the present generation, the human species, or Earth-originating life, while ignoring all other sources of value in the Universe.

First, should such an agent's background prospect have large tails? A prior question is, should it even be unbounded, or can an agent who cares only about, say, her own village be certain that its total welfare will not exceed some finite upper/lower bounds? I am inclined to say that, however narrow an agent's moral circle, 
her background prospect must be unbounded, on the grounds that we should assign probability zero only to \textit{a priori} impossibilities and, perhaps, to propositions that have been directly contradicted by observation. But this view is of course 
controversial, and I won't try to defend it here. 
Assuming that a parochial agent's background prospect should still be unbounded, however, the first and third arguments for large tails in \S \ref{section-largeTails} still seem to apply: Thinner tails would require the agent to have enormous relative confidence about very similar hypotheses that her evidence does little if anything to distinguish. 
And higher-level uncertainty 
will result in a 
heavy-tailed mixture distribution as her background prospect.


But what about dispersion? Certainly an agent who cares only about (say) her village will have a lower-dispersion background prospect than one who cares impartially about all welfare subjects in the Universe. 
Nevertheless, it seems plausible that even she will find that the dispersion of her background prospect is large relative to the possible simple payoffs of her options in most ordinary choice situations. Even my uncertainty about my own lifetime welfare, excluding the 
outcome of my present choice, is quite large (i.e., high-dispersion) relative to the stakes I face in most ordinary choice situations. And all the more so for my uncertainty about the welfare of my family, community, or nation. Finally, the third argument for high dispersion in \S \ref{section-highDispersion} seems fairly general: As long as there are many similar choices with unknown and independent effects on the things I care about, the dispersion of my uncertainty about their cumulative outcome 
is likely to be large relative to the stakes of my present choice.


\section{The rational significance of background uncertainty}
\label{section-relevanceofbackgrounduncertainty}

An initially counterintuitive feature of the preceding arguments is their implication that what an agent rationally ought to do 
can depend on her uncertainties about seemingly irrelevant features of the world. To put the point as sharply as possible: Whether I am rationally required, for instance, to take a risky action in a life-or-death situation  
can depend on my uncertainties about the existence, number, and welfare of sentient beings in distant galaxies, perhaps outside the observable universe, with whom I will never and can never interact, on whom my choices have no effect, and whose existence, number, welfare, etc, make no difference to the local effects of my choices.

Surprising and counterintuitive though this conclusion may seem, however, I think it is fully intelligible on reflection. In this section, I will try to dispel (or at least mitigate) the feeling of counterintuitiveness. 
To do that, I will first describe a simple case where the rational relevance of background uncertainty is intuitively clear, then argue that what is true of this simple case is true of more complex cases as well. 

Here,  then, is the simple case:

\begin{description}
	
	\item[\textbf{Methuselah's Choice}] Methuselah is, and knows himself to be, the only sentient being in the Universe (past, present, or future). He came into existence finitely long ago, and has so far been in a neutral state. He now faces a choice---the only choice he will ever make. He can choose either $O_1$, which yields 100 years of happy life for sure, or $O_2$, which yields 1500 years of happy life with probability $0.1$, or zero years of happy life with probability $0.9$.
		
\end{description}

If these years of happy life are the only potential source of value 
in the Universe, it seems intuitively obvious to me that Methuselah is rationally permitted to make either choice. Even if he is rationally required to satisfy the VNM axioms, say, these alone do not tell him which option to choose. And long-run arguments for expectationalism are irrelevant as well, since Methuselah knows for certain that there is no long run.

But now suppose that we add a source of background uncertainty:

\begin{description}
	\item[\text{Methuselah's Box}] In addition to Methuselah, Methuselah's universe contains a magic box, which contains a random number generator. After Methuselah makes his choice between $O_1$ and $O_2$, the box will generate a real number, from a Laplace distribution centered at zero with a scale parameter of 10,000, and open itself to reveal that number to Methuselah. In addition to the simple payoff from his choice, Methuselah will receive a number of happy life-years equal to the number generated by the box (if it is positive) or a number of unhappy life-years equal to the absolute value of that number (if it is negative).
\end{description}
(To avoid 
comparisons between happy and unhappy life-years, assume that whatever total payoff Methuselah receives, it will come in the form of exclusively happy or exclusively unhappy life-years. Thus, for instance, if he receives $+1500$ from his choice and $-2000$ from his box, he will experience 500 years of unhappy life. If he gets $+1500$ from his choice and $-200$ from his box, he will experience 1300 years of happy life.)

In virtue of Methuselah's uncertainty about the background payoff he will receive from his box, $O_2$ now stochastically dominates $O_1$ (assuming only that Methuselah regards happy life as better than unhappy life, more happy life as better, and more unhappy life as worse). 
And for precisely this reason, it now seems clear that Methuselah rationally ought to choose $O_2$. Absent the uncertainty that his box introduces, Methuselah could have reasoned his way to choosing $O_1$ on the grounds that if he chooses $O_1$, he will certainly receive at least 100 years of happy life, while if he choose $O_2$, he very probably will not. And there is no compelling defeater to this reasoning, provided that (as I claimed above) there is no compelling argument in this case for risk-neutrally maximizing expected happy life-years. But once we introduce the box, there \textit{is} a compelling defeater to the original justification for $O_1$. First, Methuselah is \textit{not} guaranteed to experience at least 100 years of happy life if he chooses $O_1$. Second, in fact, he has a \textit{better} chance of experiencing at least 100 years of happy life if he chooses $O_2$. And third, the same is true for \textit{any other possible payoff}: Whatever payoff he chooses to focus on, Methuselah has a better chance of a payoff at least that good if he choose $O_2$. Thus, Methuselah's background uncertainty gives him conclusive grounds for choosing $O_2$. 


The lesson of Methuselah's case generalizes straightforwardly to more ordinary choice situations. Suppose that Alice is 
a total hedonistic utilitarian 
and faces a choice between $O_1$, which will do an amount of good equivalent to 100 happy life-years with probability $1$, and $O_2$, which will do an amount of good equivalent to 1500 happy life-years with probability $0.1$, and do nothing with probability $0.9$. Suppose that Alice's beliefs about total welfare in the Universe, apart from the effects of her present choice, are described by a Laplace distribution centered at zero with a scale parameter of 10,000 happy-life-year-equivalents. 
Alice's situation is in every relevant respect equivalent to Methuselah's: It makes no difference, from a utilitarian standpoint, whether the welfare at stake is the agent's own, whether it belongs to a single welfare subject or to many, whether those subjects 
are near to the agent in space or time, etc. Just as in the case of Methuselah, therefore, we should conclude that (i) if there were nothing of moral significance in the Universe apart from the simple payoff of 
$O_1$ or $O_2$, then there would be no 
decisive justification for choosing $O_2$, but (ii) Alice's background uncertainty, by making $O_2$ stochastically dominant over $O_1$, gives her just such a decisive justification.

Does this 
reasoning apply only to rigorously orthodox utilitarians like Alice, who are committed to universal impartiality and the interpersonal fungibility of welfare? No. All 
it really depends on is the fungibility of \textit{choiceworthiness}, 
which is a conceptual truth so trivial 
it is hardly worth stating. Suppose that Bob accepts a commonsense, pluralistic theory of practical reasons, 
faces a choice between $O_1$ and $O_2$, where $O_1$ has a simple prospect of $\{\langle 100,1 \rangle\}$, $O_2$ has a simple prospect of $\{ \langle 0,0.9 \rangle, \langle 1500,0.1 \rangle \}$, and his background uncertainty is Laplacian with a scale parameter of 10,000. Then for any degree of choiceworthiness, $O_2$ gives Bob a better chance of performing an action at least that choiceworthy, which provides a uniquely decisive justification for choosing $O_2$. 
The contributors to choiceworthiness may be more complex and heterogeneous 
in Bob's case than in Methuselah's. But this merely obscures, and does not undermine, 
the simple argument for avoiding stochastically dominated options. 
%
%
From the standpoint of rational choice, Bob's case is no different from Methuselah's.

\section{Two modest conclusions}
\label{section-twoModestConclusions}


What decision-theoretic conclusions should we take away from the preceding arguments? In this section, I describe two relatively moderate conclusions we might draw. In the next section, I make the case for my own more ambitious conclusion.

\subsection{A decision theory for consequentialists?}

In recent years, there has been 
much activity at the intersection of ethics and decision theory, and considerable interest in the idea of `ethical decision theory'---a decision theory distinct from expected utility theory that either governs ethical decision-making in general or serves as the decision-theoretic component of particular ethical theories. 
Along these lines, the results in \S \ref{section-results} might be seen as laying the foundation for a `utilitarian decision theory', analogous to 
recent attempts to develop a `deontological decision theory' \citep{colyvan2010modelling,isaacs2014duty,lazar2017deontological}. 
Though I have argued 
that the significance of these results generalizes well beyond purely consequentialist theories like classical utilitarianism, their significance is most clear and straightforward in that context. 
For instance, the 
additive separability of 
simple and background payoffs is trivial for classical utilitarians, 
and as we saw in \S \ref{section-sourcesofbackgrounduncertainty}, uncertainty about total welfare in the Universe provides an especially strong source of background uncertainty. We might conclude from the preceding arguments, then, that SDTR is an attractive ethical decision theory for classical utilitarians and other aggregative consequentialists.

At a minimum, though, we have found that accounting for background uncertainty gives aggregative consequentialists a new and powerful basis for choosing options whose simple prospects maximize expected objective value (and not just the expectation of some increasing function of objective value) in most ordinary choice situations---even if they are also subject to decision-theoretic requirements besides stochastic dominance. That is, we have reached an important practical conclusion for aggregative consequentialists 
that requires no decision-theoretic assumptions besides the almost entirely uncontroversial SDR. \textit{A fortiori}, this conclusion applies to any aggregative consequentialist who satisfies any of the standard axiom systems like VNM or Savage, or even non-standard axiom systems like that of Buchak's (\citeyear{buchak2013risk}) REU (which, like VNM and Savage, satisfies stochastic dominance). Any such agent must, in practice, rank options almost exactly by the expectations of their simple prospects, 
even if she is extremely risk-averse or risk-seeking 
with respect to objective value (except in Pascalian situations where, as we have seen, she may enjoy greater latitude).



\subsection{An add-on to standard decision theory?}
\label{section-addOnToExpectationalism}

Building on the last observation, 
we can understand the results in \S \ref{section-results} as providing a friendly `add-on' to axiomatic expectationalism: Under sufficient background uncertainty, the standard axioms (via SDR) impose strict constraints on an agent's ranking of simple prospects, constraints that don't follow from those axioms in the absence of background uncertainty. Specifically, agents can be rationally required to rank options in a way that closely approximates the expectational ranking of their simple prospects 
\textit{under a particular, privileged assignment of cardinal values to payoffs}
---namely, the assignment that satisfies additive separability between simple and background  payoffs.\footnote{Remember that this assignment, if it exists, is unique up to positive affine transformation. So any non-affine transformation of this assignment 
will break the additive separability condition on which the results in \S \ref{section-results} depend. Perhaps more to the point, stochastic dominance relations only depend on the ordinal ranking of payoffs, so the same stochastic dominance relations will hold under a positive monotone but non-affine transformation of the privileged cardinal choiceworthiness assignment. These relations will no longer be accurately described by the Sufficiency and Necessity Theorems, however, so we cannot link stochastic dominance with expectational superiority under the transformed assignment, but only by adverting to the original, privileged assignment.} 

Plausibly, this privileged cardinalization will match the natural cardinal structure of the phenomena in the world to which the agent attaches normative weight. 
For instance, suppose that I only care about my lifetime income, always preferring more income to less. The only assignments of cardinal values to outcomes that allow additive separability between simple payoffs (the monetary reward of my present choice) and background payoffs (the remainder of my lifetime income) will be those that are positive affine transformations of the monetary value of payoffs, as measured in a currency like dollars or euros. 
So under sufficient background uncertainty, SDR and any axiomatic theory that implies it will require me to rank my options approximately by the \textit{expected monetary value} of their simple prospects. 

To put the point a little differently: Under sufficient background uncertainty, the standard axioms (by way of SDR) let us derive strong decision-theoretic conclusions merely from the agent's ranking of payoffs, 
without any information 
about her ranking of uncertain prospects. The add-on to standard decision theory here is not SDR, which was already implied, but rather the idea that agents often are or ought to be in a state of large-tailed background uncertainty. 
Recognizing this sort of background uncertainty does not impose any new constraints on the agent's utility function \textit{per se}: 
Given a ranking of overall payoffs, she may still maximize the expectation of any utility function that is increasing with respect to that ranking. But background uncertainty forces all these utility functions to agree much more than they otherwise would on the ranking of options, in a way that makes it \textit{look as if} the agent was simply maximizing her expected simple payoff on a privileged cardinal scale.\footnote{The results in \S \ref{section-results} have another, closely related implication which should be welcome news to orthodox decision theorists: They lend support to the already widely recognized idea that, if we adopt a `grand world' rather than a `small world' framing of decision problems and account for the level of background uncertainty that the grand world context implies for real-world agents, non-standard decision theories like RDU/REU are likely to end up in close practical agreement with standard decision theory. For existing arguments to this effect, see for instance \cite{quiggin2003background}, \cite{thoma2017risk} and \cite{thoma2018risk}. The existing literature tends to assume 
background uncertainty with only bounded support or thin tails, and that the agent's (non-EU-compliant) risk attitude comes from some narrowly constrained class (e.g., a transformation 
of cumulative probabilities 
of the form $f(x) = x^c$ for some constant $c$). 
But when background uncertainty is sufficient to generate stochastic dominance, 
it constrains the implications of a much wider class of attitudes toward risk, namely, all those that satisfy stochastic dominance, including all those permitted by RDU or REU.} 


As promised in \S \ref{section-exepctationalism}, 
I haven't given any novel arguments for rejecting any of the standard axioms of expected utility theory, except 
to show that we can derive 
strong and intuitively attractive practical conclusions about choices under uncertainty without appeal to those axioms. If you are inclined to accept the standard axioms, then, it is natural to adopt this `add-on' interpretation of the preceding arguments, 
as supplementing rather than supplanting 
axiomatic expectationalism.




\section{Stochastic dominance as the criterion of rational choice}
\label{section-stochasticdominancedecisiontheory}



But 
I will defend a more ambitious conclusion: that SDTR rather than expectationalism is the correct theory of rational choice under uncertainty.  
Or at least, 
I will argue that this view deserves consideration. My argument, in short, in this: The major disadvantage of SDTR relative to expectationalism is its apparent failure to place plausible constraints on our risk attitudes. 
On the other hand, SDTR has a number of advantages over expectationalism, some of which we've already seen and others of which will be described in this section. These advantages are significant enough that, if SDTR \textit{can} in fact recover intuitively satisfactory constraints on our risk attitudes in real-world choice situations, 
then it deserves to be seen as a serious competitor to expectationalism.

We have already seen two possible advantages of SDTR. First, its requirements rest on stronger \textit{a priori} foundations than those of expectationalism. 
Second, unlike primitive expectationalism, it can constrain our risk attitudes in ordinary situations while avoiding fanaticism in Pascalian situations (without recourse to ad hoc devices like excluding `\textit{de minimis} probabilities'). 
In this section, I will briefly survey some other cases where SDTR outperforms primitive and/or axiomatic expectationalism. Some of these are still problem cases for SDTR, where it is not obvious what stochastic dominance reasoning will imply or where it gives less guidance than we would like. But in all of them, SDTR delivers better answers than expectationalism seems capable of providing.

In this survey, I will mainly ignore the effects of background uncertainty. Describing those effects 
in each of the problem cases discussed below is (at least) a paper unto itself, and my aim is only to illustrate that there is a broad range of 
cases in which SDTR outperforms expectationalism.\footnote{As far as I have been able to discover, the presence of background uncertainty only ever favors SDTR (in particular, because background uncertainty can only ever generate new stochastic dominance relations among options, 
never undo existing relations \citep[p.\@ 1880]{pomatto2020stochastic}) and only ever disfavors expectationalism (in particular, by generating undefined expectations), though of course this is an imprecise and speculative claim in need of further support.} 

\subsection{Infinite payoffs}
\label{section-infinitepayoffs}


%
%

The simplest problem cases for expectational decision theory are those involving possibilities of infinite positive and/or negative payoffs, as exemplified by Pascal's Wager (\citeauthor{pascal1852pensees}, 1669). In these cases, expectational reasoning delivers either implausible advice or no advice at all. On the other hand, even in the absence of background uncertainty, stochastic dominance can often deliver plausible verdicts. To illustrate, let's consider a few variants of the Wager.

\newcounter{cases}

\vspace{2mm}

\noindent
\textbf{Case \refstepcounter{cases}\label{case-PascalsWagerCostly}\arabic{cases}: Pascal's Wager (Costly)}

\vspace{-2mm}

\begin{itemize}
	\item[$O_1$] $\{ \langle 10, 1 \rangle\}$
	
	\item[$O_2$] $\{ \langle 9, 0.99 \rangle, \langle + \infty, 0.01 \rangle \}$
	
\end{itemize}
Here, expectationalism implies that $O_2$ is rationally required, while SDTR implies that either option is rationally permissible.\footnote{I assume here (and in the discussion of infinite ethics below) that we generalize addition and multiplication, and hence expectations, in the natural way from the reals $\mathbb{R}$ to the extended reals $\mathbb{R} \cup \{\infty, - \infty \}$. In particular, $\infty x = \infty$ and $- \infty x = - \infty$ for any $x > 0$, and $\infty + (- \infty)$ is undefined. These assumptions are typical in discussions of Pascal's Wager.} 

\vspace{2mm}
\noindent
\textbf{Case \refstepcounter{cases}\label{case-PascalsWagerCostless}\arabic{cases}: Pascal's Wager (Costless)}

\vspace{-2mm}

\begin{itemize}
	\item[$O_1$] $\{ \langle 10, 1 \rangle\}$
	
	\item[$O_2$] $\{ \langle 10, 0.99 \rangle, \langle + \infty, 0.01 \rangle \}$	
\end{itemize}
Here, both SDTR and expectationalism imply that $O_2$ is rationally required.

\vspace{2mm}
\noindent
\textbf{Case \refstepcounter{cases}\label{case-PascalsWagerRegular}\arabic{cases}: Pascal's Wager (Regular)}

\vspace{-2mm}

\begin{itemize}
	\item[$O_1$] $\{ \langle 10, 0.99 \rangle, \langle + \infty, 0.01 \rangle\}$
	
	\item[$O_2$] $\{ \langle 10, 0.9 \rangle, \langle + \infty, 0.1 \rangle \}$	
\end{itemize}
Here, expectationalism implies that both options are equally good, and hence rationally permissible. SDTR implies $O_2$ is rationally required.\footnote{SDTR thus furnishes a simple reply to the `mixed strategies' objection to Pascal's Wager raised in \cite{hajek2003waging}, while also allowing that one is not \textit{always} rationally required to accept the Wager.}

\vspace{2mm}


\noindent
\textbf{Case \refstepcounter{cases}\label{case-PascalsWagerAngryGod}\arabic{cases}: Pascal's Wager (Angry God)}

\vspace{-2mm}

\begin{itemize}
	\item[$O_1$] $\{ \langle - \infty, 0.09 \rangle , \langle 9, 0.9 \rangle, \langle + \infty, 0.01 \rangle\}$
	
	\item[$O_2$] $\{ \langle - \infty, 0.01 \rangle , \langle 9, 0.9 \rangle, \langle + \infty, 0.09 \rangle\}$
\end{itemize}
Here, the expected choiceworthiness of both option is undefined, so insofar as expectationalism yields any practical conclusions at all, it implies that both options are rationally permissible. SDTR implies that $O_2$ is rationally required.\footnote{From the results in \S \ref{section-results}, we can draw 
a few conclusions about infinite payoffs under large-tailed background uncertainty. First, SDTR always permits choosing an option that increases the probability of an infinite positive payoff (or decreases the probability of an infinite negative payoff) relative to its alternatives. Second, just as with finite payoffs, large-tailed background uncertainty will sometimes generate new stochastic dominance relations between options whose simple prospects involve infinite payoffs. Among other things, this means we can put a \textit{minimum price} on Pascal's Wager. That is, if accepting the Wager increases the probability of an infinite positive payoff by $p$, then there is some finite threshold $t$ such that the Wager stochastically dominates any sure simple payoff less than $t$. If the Wager has the simple prospect $\{ \langle 0, 1 - p \rangle, \langle + \infty, p \rangle \}$, then this threshold can be expressed as $t : \min_x (\bucdf (x - t) - (1 - p) \bucdf (x)) = 0$, where $\bucdf$ is the CDF of the agent's background prospect. 
If, for instance, $p = 0.01$ and the agent's background prospect is 
Laplacian with a scale parameter of 1000, then the minimum price she is required to pay for Pascal's Wager will be slightly greater than 10.}


\subsection{The St.\@ Petersburg game}

In the St.\@ Petersburg game (\citeauthor{bernoulli1954exposition}, 1738), you are offered the chance to pay some finite price for a lottery ticket that pays $+2$ with probability $\frac{1}{2}$, $+4$ with probability $\frac{1}{4}$, $+8$ with probability $\frac{1}{8}$, and so on. 
Since the ticket has infinite expected choiceworthiness, expectationalism implies, implausibly, that you should be willing to pay any finite price for it. Once again, SDTR can do better.

\vspace{2mm}


\noindent
\textbf{Case \refstepcounter{cases}\label{case-StPetersburg}\arabic{cases}: St.\@ Petersburg}

\vspace{-2mm}

\begin{itemize}
	\item[$O_1$] $\{ \langle 100, 1 \rangle\}$
	
	\item[$O_2$] $\{ \langle 2, 0.5 \rangle , \langle 4, 0.25 \rangle, \langle 8, 0.125 \rangle, ... \}$
\end{itemize}
Here, expectationalism implies that $O_2$ is rationally required. SDTR implies that both options are rationally permissible.

\vspace{2mm}
\noindent
\textbf{Case \refstepcounter{cases}\label{case-StPetersburgPlusOne}\arabic{cases}: St.\@ Petersburg, St.\@ Petersburg +1}

\vspace{-2mm}

\begin{itemize}
	\item[$O_1$] $\{ \langle 100, 1 \rangle\}$
	
	\item[$O_2$] $\{ \langle 2, 0.5 \rangle , \langle 4, 0.25 \rangle, \langle 8, 0.125 \rangle, ... \}$

	\item[$O_3$] $\{ \langle 2 + 1, 0.5 \rangle , \langle 4 + 1, 0.25 \rangle, \langle 8 + 1, 0.125 \rangle, ... \}$

\end{itemize}
Here, expectationalism implies that $O_2$ and $O_3$ are both rationally permissible, but $O_1$ is rationally prohibited. SDTR implies that $O_1$ and $O_3$ are both rationally permissible, but $O_2$ is rationally prohibited.\footnote{As with Pascal's Wager, large-tailed background uncertainty lets us put a minimum price on the St.\@ Petersburg game, which increases under increasing rescalings of the background prospect. Under sufficient background uncertainty, clearly, the St.\@ Petersburg game can stochastically dominate arbitrarily large sure-thing payoffs, since its finite truncations can do so. The Necessity Theorem implies that it can also \textit{fail} to stochastically dominate finite sure-thing payoffs: Where $O_i$ is a St.\@ Petersburg gamble and $O_j$ yields a sure simple payoff of $t$, as $t$ goes to infinity, $\max_x \cdfDiff{i}{j}(x)$ goes to 0, while $\max_x\int_{- \infty}^{\infty} \cdfDiff{i}{j}^-(x - y) \bupdf(y) \, dy$ is non-zero and increasing. These facts together imply the existence of a minimum price.}


\subsection{The Pasadena game}

The Pasadena game \citep{nover2004vexing} is a gamble 
in which the probability-weighted sums of both positive and negative simple payoffs diverge to infinity. 
This means that the expected choiceworthiness of the gamble is not infinite but undefined.\footnote{The probability-weighted sum of possible simple payoffs of the Pasadena game is conditionally convergent, meaning that it can be made to converge to any finite value, diverge to $+/- \infty$, or simply fail to converge, depending on the order of summation. As stipulated in \S \ref{section-exepctationalism}, I assume that simple payoffs have no privileged ordering, and therefore that conditionally convergent gambles have undefined expectations. For discussion of possible extensions of expectational decision theory to handle cases like the Pasadena game, see for instance \cite{easwaran2008strong}, \cite{colyvan2008relative}, \cite{bartha2016making}, 
and \cite{lauwersvallentyne2016decision}.} In the original version of the game, we toss a fair coin until it lands heads, and receive a payoff of $(-1)^{n-1} \times \frac{2^n}{n}$, where $n$ is the number of flips. 

We can say more or less the same things about the Pasadena game as we said about the St.\@ Petersburg game.

\vspace{2mm}
\noindent
\textbf{Case \refstepcounter{cases}\label{case-Pasadena}\arabic{cases}: Pasadena}

\vspace{-2mm}

\begin{itemize}
	\item[$O_1$] $\{ \langle 100, 1 \rangle\}$
	
	\item[$O_2$] $\{ \langle 2, 0.5 \rangle , \langle -2, 0.25 \rangle, \langle \frac{8}{3}, 0.125 \rangle, -4, 0.0625 \rangle, ... \}$
\end{itemize}
Here, SDTR and expectationalism agree: The expectation of $O_2$ is undefined, and therefore incomparable with the expectation of $O_1$, so expectationalism implies that both options are permissible. SDTR straightforwardly implies that both options are permissible, since neither is stochastically dominant.

But now consider...

\vspace{2mm}
\noindent
\textbf{Case \refstepcounter{cases}\label{case-Altadena}\arabic{cases}: Pasadena, Altadena}

\vspace{-2mm}

\begin{itemize}
	\item[$O_1$] $\{ \langle 100, 1 \rangle\}$
	
	\item[$O_2$] $\{ \langle 2, 0.5 \rangle , \langle -2, 0.25 \rangle, \langle \frac{8}{3}, 0.125 \rangle, -4, 0.0625 \rangle, ... \}$

	\item[$O_3$] $\{ \langle 2 + 1, 0.5 \rangle , \langle -2 + 1, 0.25 \rangle, \langle \frac{8}{3} + 1, 0.125 \rangle, -4 + 1, 0.0625 \rangle, ... \}$
\end{itemize}
Here, since both $O_2$ and $O_3$ have undefined expectations, expectationalism implies that neither of them is comparable with $O_1$, and all three options are rationally permissible. SDTR, on the other hand, yields the intuitively correct verdict that $O_1$ and $O_3$ are rationally permissible but $O_2$ is not.\footnote{\label{footnote-lauwersVallentyneObjection}Extending an argument from \cite{seidenfeld2009preference}, \cite{lauwersvallentyne2016decision} object to 
SDR in the context of gambles without finite expectations. 
Their purported counterexample involves two anti-correlated St.\@ Petersburg lotteries, $\mathrm{SP}1$ and $\mathrm{SP}2$ (each giving its minimum payoff in exactly those states where the other does not), along with a slightly improved St.\@ Petersbury lottery $\mathrm{W}+$. Although $\mathrm{W}+$ stochastically dominates both $\mathrm{SP}1$ and $\mathrm{SP}2$, the lottery $\frac{\mathrm{SP}1 + \mathrm{SP}2}{2}$ (which yields the average of $\mathrm{SP}1$'s and $\mathrm{SP}2$'s payoff in each state) \textit{statewise} dominates (and hence stochastically dominates) $\mathrm{W}+$. SDR therefore implies that $\frac{\mathrm{SP}1 + \mathrm{SP}2}{2}$ is strictly better than both $\mathrm{SP}1$ and $\mathrm{SP}2$, which Lauwers and Vallentyne find implausible\ifanon \space (p.\@ 405). \else: `If $\mathrm{W}+$ is more valuable than $\mathrm{SP}1$ and more valuable than $\mathrm{SP}2$, then it can't be case that $(\mathrm{SP}1+\mathrm{SP}2)$/$2$ is more valuable than $\mathrm{W}+$. Thus, Stochastic Dominance must be rejected' (p.\@ 405).\fi



	
As far as I can see, though, this is simply a case of hasty generalization from finite to infinite cases. The real lesson of the example is that, when two options have infinite expectations, averaging their payoffs \textit{can} result in an improvement over both options. This is wholly plausible when we consider the result: $\mathrm{SP}1$ and $\mathrm{SP}2$ each have the simple prospect $\{ \langle 2, 0.5 \rangle , \langle 4, 0.25 \rangle, \langle 8, 0.125 \rangle, ... \}$, whereas $\frac{\mathrm{SP}1 + \mathrm{SP}2}{2}$ has the simple prospect $\{ \langle 3, 0.5 \rangle , \langle 5, 0.25 \rangle, \langle 9, 0.125 \rangle, ... \}$. The result of averaging the two anti-correlated St.\@ Petersburg lotteries, in other words, is St.\@ Petersburg +1. As long as we accept that this is an improvement over St.\@ Petersburg, 
this case gives us no reason to question SDR. 
(For another, more thorough 
reply to this challenge to SDR, 
see \cite{meacham2019difference}.)}


\subsection{Ordinality and lexicality}
\label{section-ordinal}

Philosophers have recently begun paying attention to decision-theoretic questions that arise when an agent is uncertain not only about the empirical state of the world but also about basic normative principles.\footnote{
For a survey of this literature, see \cite{bykvist2017moral}.} As has been noted in this literature (e.g.\@ by \cite{sepielli2010along} and \cite{macaskill2016normative}), a major difficulty for extending standard expectational decision theory to this `metanormative' context is that some normative theories appear to give only ordinal rankings of options, which cannot be multiplied by probabilities to compute expectations and which expectational reasoning is therefore unable to handle. This has led to the suggestion (e.g.\@ in \cite{macaskill2014normative}) that fundamentally different decision procedures may be needed to handle different categories of normative theory. Stochastic dominance reasoning, however, can handle both ordinal and cardinal contexts, and may therefore offer a more unified theory of rational choice than expectationalism, given the existence of merely-ordinal normative theories. As a simple illustration, consider the following case, where Roman numerals represent ordinal ranks, with larger numerals representing greater degrees of choiceworthiness.


\vspace{2mm}


\noindent
\textbf{Case \refstepcounter{cases}\label{case-OrdinalRisk}\arabic{cases}: Ordinal Risk}

\vspace{-2mm}

\begin{itemize}
	\item[$O_1$] $\{ \langle i, 0.4 \rangle, \langle ii, 0.6 \rangle\}$
	
	\item[$O_2$] $\{ \langle ii, 0.7 \rangle, \langle iii, 0.3 \rangle\}$
	
\end{itemize} 
Since the payoffs have only ordinal values, the expected choiceworthiness of both options is of course undefined. 
But SDTR correctly implies that $O_2$ is rationally required.

Another worry in the literature on normative uncertainty is that some normative theories rank options \textit{lexically}, either regarding certain categories of action as absolutely required or prohibited (e.g., 
punishing the innocent), or regarding certain categories of normative consideration as taking absolute precedence over others (e.g., the welfare of the worse off over the welfare of the better off). It is not obvious how these theories should should be represented for decision-theoretic purposes (for discussion, see \cite{colyvan2010modelling}). But at least on a simple representation, choice situations involving such lexical considerations will have the same structure as the `infinite payoff' cases described in \S \ref{section-infinitepayoffs}, and so seem to favor SDTR over expectationalism.\ifanon\else\footnote{For more on stochastic dominance reasoning in the context of uncertainty among merely-ordinal and/or lexical normative theories, see \cite{tarsney2018moral} and \citeauthor{aboodimsworking} (unpublished).}\fi 

\subsection{Incomparability and incompleteness}

Another kind of problem case for expectationalism involves incomplete rankings of payoffs that arise when competing normative considerations are incomparable or only roughly comparable. 
As with ordinality and lexicality, the decision-theoretic problems associated with incompleteness are especially acute in the context of 
normative uncertainty: As \cite{macaskill2013infectiousness} points out, an agent who has \textit{any} positive credence in theories that posit incomparability between the possible payoffs of her options is likely to find that the expected choiceworthiness of those options is undefined. 

Once again, however, SDTR can deliver intuitive verdicts in cases of incomparability that expectational reasoning cannot. For instance, as \cite{bader2018stochastic} points out, stochastic dominance reasoning straightforwardly resolves the `opaque sweetening' case introduced by \cite{hare2010take}. Here, $a$ and $b$ represent incomparable payoffs, and $a^+$ and $b^+$ are improved versions of those payoffs, such that $a^+$ is preferable to $a$ but incomparable with $b$ and $b^+$ (and likewise $b^+$ is preferable to $b$ but incomparable with $a$ and $a^+$).

\vspace{2mm}

\noindent
\textbf{Case \refstepcounter{cases}\label{case-OpaqueSweetening}\arabic{cases}: Opaque Sweetening}

\vspace{-2mm}

\begin{itemize}
	\item[$O_1$] $\{ \langle a, 0.5 \rangle, \langle b, 0.5 \rangle\}$
	
	\item[$O_2$] $\{ \langle a^{+}, 0.5 \rangle, \langle b^{+}, 0.5 \rangle\}$
	
\end{itemize}
Once again, the expected choiceworthiness of both options is undefined, so expectationalism implies that both options are rationally permissible. SDTR more plausibly implies that $O_2$ is rationally required.\footnote{\label{footnote-opaqueSweeteningObjection}A crucial feature of Hare's case is that $O_1$ yields $a$ in the state 
where $O_2$ yields $b^{+}$, and $b$ in the state 
where $O_2$ yields $a^{+}$. \cite{bales2014decision} and \cite{schoenfield2014decision} have both denied that $O_2$ is rationally required in this case, invoking principles to the effect that, if $O_2$ is not better than $O_1$ in any possible state of nature, one is not rationally required to prefer it. Both thereby implicitly deny SDR. 
I find these arguments unpersuasive, because I don't see why the way we 
group possible world histories into `states of nature' should have any practical significance. (The argument for the permissibility of $O_1$ depends on treating the history where you choose $O_1$ and receive $a$ as belonging to the same `state of nature' as the history where you choose $O_2$ and receive $b^{+}$, and likewise for `choose $O_1$, receive $b$' and `choose $O_2$, receive $a^{+}$'. If we instead grouped `choose $O_1$, receive $a$' with `choose $O_2$, receive $a^+$' and `choose $O_1$, receive $b$' with `choose $O_2$, receive $b^+$', then statewise reasoning would imply that $O_2$ is rationally required. But fundamentally, these are just four different ways the world could be, and while grouping them in particular ways may sometimes be natural and convenient, our normative judgments should not depend on it.) But this turns on basic questions in decision theory, which I won't try to resolve here.}

\subsection{Infinite worlds}
\label{section-infiniteworlds}

%
%

As I admitted in \S \ref{section-largeTails}, the possibility that the Universe is infinite, and contains infinitely much positive and negative value \textit{regardless} of our choices, complicates my central line of argument. Generalizing the results and arguments in \S \S \ref{section-results}--\ref{section-sourcesofbackgrounduncertainty} to this context
requires a satisfactory axiology for infinite worlds, which we don't yet have. But at least on face, SDTR seems much better equipped than expectationalism to handle infinite worlds. The simplest axiological representation of infinite worlds is given by the extended real number line (the reals, plus special elements $\infty$ and $-\infty$, ordered as you would expect). This is the worst case for consequentialist ethical reasoning, since it implies that no finite difference we can make to the world has any axiological effect. Nonetheless, even under this gloomy supposition, SDTR is able to provide useful practical guidance, so long as the agent has non-zero credence that the world is finite. Suppose, for instance, that she is nearly certain that the world is infinite and contains either infinite positive value or infinite negative value, but has some credence that it is finite, such that her actions can make an axiological difference.

\vspace{2mm}
\noindent
\textbf{Case \refstepcounter{cases}\label{case-InfiniteWorldsHeavenOrHell}\arabic{cases}: Heaven or Hell}

\vspace{-2mm}

\begin{itemize}
	\item[$O_1$] $\{ \langle - \infty, 0.45 \rangle , \langle 10, 0.1 \rangle, \langle + \infty, 0.45 \rangle\}$
	
	\item[$O_2$] $\{ \langle - \infty, 0.45 \rangle , \langle 11, 0.1 \rangle, \langle + \infty, 0.45 \rangle\}$
\end{itemize}
Here, the expected choiceworthiness of both options is undefined, so expectationalism implies that both options are rationally permissible, while SDTR correctly implies that $O_2$ is rationally required.

This is just a simple illustration of a broader point: If $O_i$ and $O_j$ each carry the same probabilities of infinite positive and infinite negative payoffs, then $O_i$ stochastically dominates $O_j$ just in case its \textit{finite} prospect is stochastically dominant. Thus, if we can't change the probabilities of infinite payoffs, SDTR (unlike expectationalism) allows us to simply ignore the infinite possibilities and condition our choice on the assumption of a finite payoff. In this way at least, the positive features of SDTR under background uncertainty established in \S \ref{section-results} transfer straightforwardly to the infinite context.

Things get slightly tricker when we consider the more realistic possibility that the world, being infinite, contains infinitely much of both positive \textit{and} negative value. Here it is not only the expectation but the cardinal value itself that is undefined. However, if we are willing to treat $\infty - \infty$ as a special degree of value, albeit one that is incomparable with any 
finite 
degree of value, then the same conclusions will hold:

\vspace{2mm}


\noindent
\textbf{Case \refstepcounter{cases}\label{case-InfiniteWorldsHeavenAndHell}\arabic{cases}: Heaven + Hell}

\vspace{-2mm}

\begin{itemize}
	\item[$O_1$] $\{ \langle - \infty, 0.05 \rangle , \langle 10, 0.1 \rangle, \langle + \infty, 0.05 \rangle, \langle \infty - \infty, 0.8 \rangle\}$
	
	\item[$O_2$] $\{ \langle - \infty, 0.05 \rangle , \langle 11, 0.1 \rangle, \langle + \infty, 0.05 \rangle, \langle \infty - \infty, 0.8 \rangle\}$
\end{itemize}
Here again, expectationalism is silent, while SDTR implies that $O_2$ is rationally required: Since $10 < 11$, and both options give the same probability of every other simple payoff, $O_2$ stochastically dominates $O_1$.
\footnote{One might be tempted to think that we should treat the probability assigned to $\infty - \infty$ like pure Knightian uncertainty over the whole extended real number line, in which case we could not say, for instance, that $O_2$ offers a greater probability of a payoff at least as good as 11. But this would be a mistake: I am not uncertain whether $\infty - \infty$ is greater than, less than, or equal to 11. Rather, I am certain that the two payoffs are incomparable.} 

Of course, the extended real number line gives a supremely unsatisfying representation of the value of infinite worlds, 
and much ink has been spilled trying to do better (see note \ref{footnote-infiniteethicscites}
). I won't try to survey these accounts or describe how stochastic dominance might interact with each of them. But I will point out that, if the correct axiology allows us to make ordinal comparisons between infinite worlds, then SDTR can derive practical conclusions from uncertainty over those ordinal values. And if the correct axiology lets us make ordinal but not cardinal comparisons (between some or all pairs of infinite worlds), then SDTR is here too at an advantage over expectationalism.\footnote{Cardinal comparisons have been treated as a desideratum in the infinite ethics literature largely in order to accommodate expectational decision theory (e.g.\@ in \citeauthor{bostrom2011infinite} (\citeyear{bostrom2011infinite}, pp.\@ 21--22) and \citeauthor{arntzenius2014utilitarianism} (\citeyear{arntzenius2014utilitarianism}, p.\@ 37)). If the correct decision theory does not require cardinality, therefore, this might make it easier to find a satisfactory axiology for infinite worlds.} 

Consider, for instance, a modified version of the ordinal case from \S \ref{section-ordinal}, with Roman numeral subscripts now representing 
ordinal ranks assigned to infinite worlds.

\vspace{2mm}
\noindent
\textbf{Case \refstepcounter{cases}\label{case-InfiniteWorldsOrdinalHeavenAndHell}\arabic{cases}: Ordinal Heaven + Hell}

\vspace{-2mm}

\begin{itemize}
	\item[$O_1$] $\{ \langle 10, 0.2 \rangle, \langle 15, 0.3 \rangle, \langle (\infty - \infty)_i, 0.2 \rangle, \langle (\infty - \infty)_{ii}, 0.3 \rangle\}$
	
	\item[$O_2$] $\{ \langle 15, 0.35 \rangle, \langle 20, 0.15 \rangle, \langle (\infty - \infty)_{ii}, 0.35 \rangle, \langle (\infty - \infty)_{iii}, 0.15 \rangle\}$
	
\end{itemize}
Once again, expectationalism is silent, while SDTR correctly implies that $O_2$ is rationally required.



\section{Conclusion}
\label{section-conclusion}

Under levels of background uncertainty that are plausibly required of real-world agents, 
stochastic dominance can effectively constrain risk attitudes, recovering many of the plausible implications of expectational reasoning, 
while to a significant extent avoiding the threat of Pascalian fanaticism. These facts have important practical implications for real-world choices, giving us stronger justification for maximizing the expectation of objective goods (like lives saved, or total welfare) in most ordinary choice situations, while suggesting an escape route from the most extreme demands of risk-neutral expectational reasoning. They may also have theoretical implications for decision theory. Since stochastic dominance reasoning handles a range of problem cases better than expectational reasoning, and rests on stronger \textit{a priori} foundations, if it can also provide satisfactory constraints on real-world choices, 
then SDTR deserves consideration as a general criterion of rational choice under uncertainty.

\appendix

\section{Proofs of theorems}

\SufficiencyTheorem*

\begin{proof}
	
	
	Consider an arbitrary payoff $\arbitrarypayoff$. Given a background payoff $x$, option $O_i$ yields a payoff $\geq \arbitrarypayoff$ iff it yields a simple payoff $\geq \arbitrarypayoff - x$. Therefore, where $\Pr(O_i \geq x)$ is the probability that $O_i$ yields a simple payoff $\geq x$, the total probability that $O_i$ yields an overall payoff $\geq \arbitrarypayoff$ is given by
	
	\begin{equation}
	\ccdf{\bucdf_i}(\arbitrarypayoff) = \int_{-\infty}^{\infty} \bupdf(x) \Pr(O_i \geq \arbitrarypayoff - x) \, dx.
	\end{equation}
		
	Therefore the \textit{difference} between the probability that $O_i$ yields a payoff $\geq \arbitrarypayoff$ and the probability that $O_j$ yields a payoff $\geq \arbitrarypayoff$ is given by
	
	\begin{align}
		\ccdf{\bucdf_i}(\arbitrarypayoff) - \ccdf{\bucdf_j}(\arbitrarypayoff) & = \int_{-\infty}^{\infty} \bupdf(x) \Pr(O_i \geq \arbitrarypayoff - x) \, dx - \int_{-\infty}^{\infty} \bupdf(x) \Pr(O_j \geq \arbitrarypayoff - x) \, dx \\
		& = \int_{-\infty}^{\infty} \bupdf(x) (\Pr(O_i \geq \arbitrarypayoff - x) - \Pr(O_j \geq \arbitrarypayoff - x)) \, dx \\
		& = \int_{-\infty}^{\infty} \bupdf(x) \cdfDiff{i}{j}(\arbitrarypayoff - x) \, dx \\
		& = \int_{-\infty}^{\infty} \bupdf(x) \cdfDiff{i}{j}^+(\arbitrarypayoff - x) \, dx - \int_{-\infty}^{\infty} \bupdf(x) \cdfDiff{i}{j}^-(\arbitrarypayoff - x) \, dx.
	\end{align}
		
	
	Since $\ccdf{\bucdf_i}$, $\ccdf{\bucdf_j}$, $\bupdf$, $\cdfDiff{i}{j}^+$, and $\cdfDiff{i}{j}^-$ are all non-negative, it follows that
	
	\begin{equation}
	\ccdf{\bucdf_i}(\arbitrarypayoff) > \ccdf{\bucdf_j}(\arbitrarypayoff) \Leftrightarrow \frac{\int_{-\infty}^{\infty} \bupdf(x) \cdfDiff{i}{j}^+(\arbitrarypayoff - x) \, dx}{\int_{-\infty}^{\infty} \bupdf(x) \cdfDiff{i}{j}^-(\arbitrarypayoff - x) \, dx} > 1.\label{equation-conditionForStrongSD}
	\end{equation}
	
	By definition, the value of $\bupdf$ cannot vary over the support of $\cdfDiff{i}{j}$ by more than a factor of $\maxRatioParam(O_i,O_j,\bupdf)$. This means that
		
	
	\begin{equation}
	\frac{\int_{-\infty}^{\infty} \bupdf(x) \cdfDiff{i}{j}^+(\arbitrarypayoff - x) \, dx}{\int_{-\infty}^{\infty} \bupdf(x) \cdfDiff{i}{j}^-(\arbitrarypayoff - x) \, dx} \geq \frac{\int_{-\infty}^{\infty} \cdfDiff{i}{j}^+(\arbitrarypayoff - x) \, dx}{\maxRatioParam(O_i,O_j,\bupdf) \int_{-\infty}^{\infty} \cdfDiff{i}{j}^-(\arbitrarypayoff - x) \, dx}.\label{equation-boundingByRate}
	\end{equation}
	
	
	From lines \ref{equation-conditionForStrongSD} and \ref{equation-boundingByRate}, it follows that:
	
	\begin{equation}	
	\frac{\int_{-\infty}^{\infty} \cdfDiff{i}{j}^+(x) \, dx}{\int_{-\infty}^{\infty} \cdfDiff{i}{j}^-(x) \, dx} > \maxRatioParam(O_i,O_j,\bupdf) \Rightarrow 
	\ccdf{\bucdf_i}(\arbitrarypayoff) > \ccdf{\bucdf_j}(\arbitrarypayoff).
	\end{equation}
	
	And since this is true for arbitrary $\arbitrarypayoff$, we can conclude that
	
	\begin{equation}
	\frac{\int_{-\infty}^{\infty} \cdfDiff{i}{j}^+(x) \, dx}{\int_{-\infty}^{\infty} \cdfDiff{i}{j}^-(x) \, dx} > \maxRatioParam(O_i,O_j,\bupdf) \Rightarrow \forall x (\ccdf{\bucdf_i}(x) > \ccdf{\bucdf_j}(x)),
	\end{equation}
	which implies
	
	$$\frac{\int_{-\infty}^{\infty} \cdfDiff{i}{j}^+(x) \, dx}{\int_{-\infty}^{\infty} \cdfDiff{i}{j}^-(x) \, dx} > \maxRatioParam(O_i,O_j,\bupdf) \Rightarrow \sdArgs{O_i}{O_j}.$$
	
\end{proof}

\NecessityTheorem*

\begin{proof}
	
	From the proof of the Sufficiency Theorem, we know that $\sdArgs{O_i}{O_j}$ only if
	
	\begin{equation}
	\forall x \left(\int_{-\infty}^{\infty} \bupdf(y) \cdfDiff{i}{j}^+(x - y) \, dy \geq \int_{-\infty}^{\infty} \bupdf(y) \cdfDiff{i}{j}^-(x - y) \, dy\right).\label{equation-LbtPrfLn1}
	\end{equation}
	
	Suppose that $\cdfDiff{i}{j}$ was a constant function, with $\cdfDiff{i}{j}(x) = k$ for all $x$. Then, since $\int_{-\infty}^{\infty} \bupdf(y) \, dy = 1$, it would follow that $\int_{-\infty}^{\infty} \bupdf(y) \cdfDiff{i}{j}^+(x - y) \, dy = k$. From this we can infer that
	
	\begin{equation}
	\forall x \left(\max_z \cdfDiff{i}{j}(z) \geq \int_{-\infty}^{\infty} \bupdf(y) \cdfDiff{i}{j}^+(x - y) \, dy\right).
	\end{equation}
	
	And in fact, given that the simple prospects of $O_i$ and $O_j$ (i) are non-identical (a necessary condition for stochastic dominance) and (ii) involve only finite payoffs (as stipulated in \S \ref{section-formalSetup}), $\cdfDiff{i}{j}$ cannot be constant, so the inequality is strict: $\max_z \cdfDiff{i}{j}(z) > \int_{-\infty}^{\infty} \bupdf(y) \cdfDiff{i}{j}^+(x - y) \, dy$.
	
	From this it follows (by substitution in line \ref{equation-LbtPrfLn1}) that:
	
	\begin{equation}
	\sdArgs{O_i}{O_j} \Rightarrow \forall x \left(\max_z \cdfDiff{i}{j}(z) > \int_{-\infty}^{\infty} \bupdf(y) \cdfDiff{i}{j}^-(x - y) \, dy\right),
	\end{equation}
or in other words
	\begin{equation*}
	\sdArgs{O_i}{O_j} \Rightarrow \max_x \cdfDiff{i}{j}(x) > \max_x \int_{-\infty}^{\infty} \cdfDiff{i}{j}^-(x - y) \bupdf(y) \, dy.
	\end{equation*}
	
	
%
	
\end{proof}

\bibliography{stochasticdominancecites}

\begin{thebibliography}{74}
\providecommand{\natexlab}[1]{#1}
\providecommand{\url}[1]{\texttt{#1}}
\expandafter\ifx\csname urlstyle\endcsname\relax
  \providecommand{\doi}[1]{doi: #1}\else
  \providecommand{\doi}{doi: \begingroup \urlstyle{rm}\Url}\fi

\bibitem[Aboodi()]{aboodimsworking}
Ron Aboodi.
\newblock Working with ordinal intertheoretic choice-worthiness comparisons.
\newblock Unpublished ms.

\bibitem[Arntzenius(2014)]{arntzenius2014utilitarianism}
Frank Arntzenius.
\newblock Utilitarianism, decision theory and eternity.
\newblock \emph{Philosophical Perspectives}, 28\penalty0 (1):\penalty0 31--58,
  2014.

\bibitem[Bader(2018)]{bader2018stochastic}
Ralf~M. Bader.
\newblock Stochastic dominance and opaque sweetening.
\newblock \emph{Australasian Journal of Philosophy}, 96\penalty0 (3):\penalty0
  498--507, 2018.

\bibitem[Bales et~al.(2014)Bales, Cohen, and Handfield]{bales2014decision}
Adam Bales, Daniel Cohen, and Toby Handfield.
\newblock Decision theory for agents with incomplete preferences.
\newblock \emph{Australasian Journal of Philosophy}, 92\penalty0 (3):\penalty0
  453--70, 2014.

\bibitem[Bartha(2016)]{bartha2016making}
Paul F.~A. Bartha.
\newblock Making do without expectations.
\newblock \emph{Mind}, 125\penalty0 (499):\penalty0 799--827, 2016.

\bibitem[Beckstead(2013)]{beckstead2013overwhelming}
Nick Beckstead.
\newblock \emph{On the Overwhelming Importance of Shaping the Far Future}.
\newblock PhD thesis, Rutgers University Graduate School - New Brunswick, 2013.

\bibitem[Beckstead(2019)]{beckstead2019brief}
Nick Beckstead.
\newblock A brief argument for the overwhelming importance of shaping the far
  future.
\newblock In Hilary Greaves and Theron Pummer, editors, \emph{Effective
  Altruism: Philosophical Issues}, pages 80--98. Oxford University Press,
  Oxford, 2019.

\bibitem[Bernoulli(1954 (1738))]{bernoulli1954exposition}
Daniel Bernoulli.
\newblock Exposition of a new theory on the measurement of risk.
\newblock \emph{Econometrica: Journal of the Econometric Society}, 22\penalty0
  (1):\penalty0 23--36, 1954 (1738).

\bibitem[Bostrom(2009)]{bostrom2009pascal}
Nick Bostrom.
\newblock Pascal's mugging.
\newblock \emph{Analysis}, 69\penalty0 (3):\penalty0 443--445, 2009.

\bibitem[Bostrom(2011)]{bostrom2011infinite}
Nick Bostrom.
\newblock Infinite ethics.
\newblock \emph{Analysis and Metaphysics}, 10:\penalty0 9--59, 2011.

\bibitem[Bostrom(2013)]{bostrom2013existential}
Nick Bostrom.
\newblock Existential risk prevention as global priority.
\newblock \emph{Global Policy}, 4\penalty0 (1):\penalty0 15--31, 2013.

\bibitem[Briggs(2017)]{briggs2017normative}
Rachael Briggs.
\newblock Normative theories of rational choice: Expected utility.
\newblock In Edward~N. Zalta, editor, \emph{The Stanford Encyclopedia of
  Philosophy}. Metaphysics Research Lab, Stanford University, 2017.

\bibitem[Buchak(2013)]{buchak2013risk}
Lara Buchak.
\newblock \emph{Risk and Rationality}.
\newblock Oxford University Press, Oxford, 2013.

\bibitem[Buffon(2010 (1777))]{buffon1777essai}
Georges-Louis Leclerc~de Buffon.
\newblock Essai d'arithmetique morale ({E}ssays on moral arithmetic).
\newblock In Axel Ockenfels and Abdolkarim Sadrieh, editors, \emph{The Selten
  School of Behavioral Economics: A Collection of Essays in Honor of Reinhard
  Selten}, pages 245--282. Springer, Heidelberg, 2010 (1777).
\newblock Translated by John D.\@ Hey, Tibor M.\@ Neugebauer, and Carmen M.\@
  Pasca.

\bibitem[Bykvist(2017)]{bykvist2017moral}
Krister Bykvist.
\newblock Moral uncertainty.
\newblock \emph{Philosophy Compass}, 12\penalty0 (3):\penalty0 e12408, 2017.

\bibitem[{Carroll}(2017)]{carroll2017why}
Sean~M. {Carroll}.
\newblock Why {B}oltzmann brains are bad.
\newblock \emph{arXiv e-prints}, art. arXiv:1702.00850, February 2017.

\bibitem[Colyvan(2008)]{colyvan2008relative}
Mark Colyvan.
\newblock Relative expectation theory.
\newblock \emph{Journal of Philosophy}, 105\penalty0 (1):\penalty0 37--44,
  2008.

\bibitem[Colyvan et~al.(2010)Colyvan, Cox, and Steele]{colyvan2010modelling}
Mark Colyvan, Damian Cox, and Katie Steele.
\newblock Modelling the moral dimension of decisions.
\newblock \emph{No\^us}, 44\penalty0 (3):\penalty0 503--529, 2010.

\bibitem[Easwaran(2008)]{easwaran2008strong}
Kenny Easwaran.
\newblock Strong and weak expectations.
\newblock \emph{Mind}, 117\penalty0 (467):\penalty0 633--641, 2008.

\bibitem[Easwaran(2014)]{easwaran2014decision}
Kenny Easwaran.
\newblock Decision theory without representation theorems.
\newblock \emph{Philosopher's Imprint}, 14\penalty0 (27):\penalty0 1--30, 2014.

\bibitem[Feller(1968)]{feller1968introduction}
William Feller.
\newblock \emph{An Introduction to Probability Theory and Its Applications},
  volume~1.
\newblock Wiley, New York, 3rd edition, 1968.

\bibitem[Gong et~al.(2011)Gong, Zhu, and Zhu]{gong2011current}
Yungui Gong, Xiao-ming Zhu, and Zong-Hong Zhu.
\newblock Current cosmological constraints on the curvature, dark energy and
  modified gravity.
\newblock \emph{Monthly Notices of the Royal Astronomical Society},
  415\penalty0 (2):\penalty0 1943--1949, 2011.

\bibitem[Greaves and MacAskill(2019)]{greavesMScase}
Hilary Greaves and William MacAskill.
\newblock The case for strong longtermism.
\newblock \emph{Global Priorities Institute Working Paper Series}, 2019.
\newblock GPI Working Paper No.\@ 7-2019.

\bibitem[Greene(2004)]{greene2004fabric}
Brian Greene.
\newblock \emph{The Fabric of the Cosmos: Space, Time and the Texture of
  Reality}.
\newblock Random House, New York, 2004.

\bibitem[Gronwall(1919)]{gronwall1919note}
Thomas~Hakon Gronwall.
\newblock Note on the derivatives with respect to a parameter of the solutions
  of a system of differential equations.
\newblock \emph{Annals of Mathematics}, 20\penalty0 (4):\penalty0 292--296,
  1919.

\bibitem[H\'ajek(2003)]{hajek2003waging}
Alan H\'ajek.
\newblock Waging war on {P}ascal's {W}ager.
\newblock \emph{Philosophical Review}, 112\penalty0 (1):\penalty0 27--56, 2003.

\bibitem[Hare(2010)]{hare2010take}
Caspar Hare.
\newblock Take the sugar.
\newblock \emph{Analysis}, 70\penalty0 (2):\penalty0 237--247, 2010.

\bibitem[Harsanyi(1955)]{harsanyi1955cardinal}
John~C.\@ Harsanyi.
\newblock Cardinal welfare, individualistic ethics, and interpersonal
  comparisons of utility.
\newblock \emph{Journal of Political Economy}, 63\penalty0 (4):\penalty0
  309--321, 1955.

\bibitem[Isaacs(2014)]{isaacs2014duty}
Yoaav Isaacs.
\newblock Duty and knowledge.
\newblock \emph{Philosophical Perspectives}, 28\penalty0 (1):\penalty0 95--110,
  2014.

\bibitem[Jimenez et~al.(2018)Jimenez, Raccanelli, Verde, and
  Matarrese]{jimenez2018peering}
Raul Jimenez, Alvise Raccanelli, Licia Verde, and Sabino Matarrese.
\newblock Peering beyond the horizon with standard sirens and redshift drift.
\newblock \emph{Journal of Cosmology and Astroparticle Physics}, 04\penalty0
  (002):\penalty0 1--13, 2018.

\bibitem[Joyce(1999)]{joyce1999foundations}
James~M.\@ Joyce.
\newblock \emph{The Foundations of Causal Decision Theory}.
\newblock Cambridge University Press, Cambridge, 1999.

\bibitem[Kaneda and Haub(2018)]{kaneda2011how}
Toshiko Kaneda and Carl Haub.
\newblock How many people have ever lived on earth?
\newblock Population Reference Bureau, 2018.
\newblock First published in 1997, updated in 2002, 2011, and 2018. Accessed 22
  November 2019. URL: https://www.prb.org/howmanypeoplehaveeverlivedonearth/.

\bibitem[Knight(1921)]{knight1921risk}
Frank~H.\@ Knight.
\newblock \emph{Risk, Uncertainty and Profit}.
\newblock Houghton Mifflin Company, Boston, 1921.

\bibitem[Knobe et~al.(2006)Knobe, Olum, and Vilenkin]{knobe2006philosophical}
Joshua Knobe, Ken~D. Olum, and Alexander Vilenkin.
\newblock Philosophical implications of inflationary cosmology.
\newblock \emph{The British Journal for the Philosophy of Science}, 57\penalty0
  (1):\penalty0 47--67, 2006.

\bibitem[Krantz et~al.(1971)Krantz, Luce, Suppes, and
  Tversky]{krantz1971foundations}
David~H.\@ Krantz, R.\@~Duncan Luce, Patrick Suppes, and Amos Tversky.
\newblock \emph{Foundations of Measurement, Volume I: Additive and Polynomial
  Representations}.
\newblock Academic Press, New York, 1971.

\bibitem[Lauwers and Vallentyne(2016)]{lauwersvallentyne2016decision}
Luc Lauwers and Peter Vallentyne.
\newblock Decision theory without finite standard expected value.
\newblock \emph{Economics and Philosophy}, 32\penalty0 (3):\penalty0 383--407,
  2016.

\bibitem[Lazar(2017)]{lazar2017deontological}
Seth Lazar.
\newblock Deontological decision theory and agent-centered options.
\newblock \emph{Ethics}, 127\penalty0 (3):\penalty0 579--609, 2017.

\bibitem[Levy(2016)]{levy2016stochastic}
Haim Levy.
\newblock \emph{Stochastic Dominance: Investment Decision Making under
  Uncertainty}.
\newblock Springer, Cham, 3rd edition, 2016.

\bibitem[MacAskill(2013)]{macaskill2013infectiousness}
William MacAskill.
\newblock The infectiousness of nihilism.
\newblock \emph{Ethics}, 123\penalty0 (3):\penalty0 508--520, 2013.

\bibitem[MacAskill(2014)]{macaskill2014normative}
William MacAskill.
\newblock \emph{Normative Uncertainty}.
\newblock PhD thesis, University of Oxford, 2014.

\bibitem[MacAskill(2016)]{macaskill2016normative}
William MacAskill.
\newblock Normative uncertainty as a voting problem.
\newblock \emph{Mind}, 125\penalty0 (500):\penalty0 967--1004, 2016.

\bibitem[MacAskill and Ord(2020)]{macaskill2020why}
William MacAskill and Toby Ord.
\newblock Why maximize expected choice-worthiness?
\newblock \emph{No\^us}, 54\penalty0 (2):\penalty0 327--353, 2020.

\bibitem[Manski(2011)]{manski2011actualist}
Charles~F. Manski.
\newblock Actualist rationality.
\newblock \emph{Theory and Decision}, 71\penalty0 (2):\penalty0 195--210, 2011.

\bibitem[Meacham(2019)]{meacham2019difference}
Christopher J.~G. Meacham.
\newblock Difference minimizing theory.
\newblock \emph{Ergo: An Open Access Journal of Philosophy}, 6\penalty0
  (35):\penalty0 999--1034, 2019.

\bibitem[Meacham and Weisberg(2011)]{meacham2011representation}
Christopher J.~G. Meacham and Jonathan Weisberg.
\newblock Representation theorems and the foundations of decision theory.
\newblock \emph{Australasian Journal of Philosophy}, 89\penalty0 (4):\penalty0
  641--663, 2011.

\bibitem[Mulgan(2002)]{mulgan2002transcending}
Tim Mulgan.
\newblock Transcending the infinite utility debate.
\newblock \emph{Australasian Journal of Philosophy}, 80\penalty0 (2):\penalty0
  164--177, 2002.

\bibitem[Ng(1997)]{ng1997case}
Yew-Kwang Ng.
\newblock A case for happiness, cardinalism, and interpersonal comparability.
\newblock \emph{The Economic Journal}, 107\penalty0 (445):\penalty0 1848--1858,
  1997.

\bibitem[Nover and H\'ajek(2004)]{nover2004vexing}
Harris Nover and Alan H\'ajek.
\newblock Vexing expectations.
\newblock \emph{Mind}, 113\penalty0 (450):\penalty0 237--249, 2004.

\bibitem[Ord(2020)]{ord2020precipice}
Toby Ord.
\newblock \emph{The Precipice: Existential Risk and the Future of Humanity}.
\newblock Bloomsbury Publishing, London, 2020.

\bibitem[Page(2007)]{page2007susskinds}
Don~N.\@ Page.
\newblock Susskind's challenge to the {H}artle-{H}awking no-boundary proposal
  and possible resolutions.
\newblock \emph{Journal of Cosmology and Astroparticle Physics}, 01\penalty0
  (004):\penalty0 1--20, 2007.

\bibitem[Parfit(2011)]{parfit2011onwhatmatters}
Derek Parfit.
\newblock \emph{On What Matters}.
\newblock Oxford University Press, Oxford, 2011.

\bibitem[Pascal(1852 (1669))]{pascal1852pensees}
Blaise Pascal.
\newblock \emph{Pens{\'e}es}.
\newblock Dezobry et E. Magdeleine, Paris, 1852 (1669).

\bibitem[Pomatto et~al.(2020)Pomatto, Strack, and Tamuz]{pomatto2020stochastic}
Luciano Pomatto, Philipp Strack, and Omer Tamuz.
\newblock Stochastic dominance under independent noise.
\newblock \emph{Journal of Political Economy}, 128\penalty0 (5):\penalty0
  1877--1900, 2020.

\bibitem[Quiggin(1982)]{quiggin1982theory}
John Quiggin.
\newblock A theory of anticipated utility.
\newblock \emph{Journal of Economic Behavior \& Organization}, 3\penalty0
  (4):\penalty0 323--343, 1982.

\bibitem[Quiggin(2003)]{quiggin2003background}
John Quiggin.
\newblock Background risk in generalized expected utility theory.
\newblock \emph{Economic Theory}, 22\penalty0 (3):\penalty0 607--611, 2003.

\bibitem[{Sandberg} et~al.(2018){Sandberg}, {Drexler}, and
  {Ord}]{sandberg2018dissolving}
Anders {Sandberg}, Eric {Drexler}, and Toby {Ord}.
\newblock Dissolving the fermi paradox.
\newblock \emph{arXiv e-prints}, art. arXiv:1806.02404, June 2018.

\bibitem[Savage(1954)]{savage1954foundations}
Leonard~J. Savage.
\newblock \emph{The Foundations of Statistics}.
\newblock John Wiley \& Sons, New York, 1st edition, 1954.

\bibitem[Schoenfield(2014)]{schoenfield2014decision}
Miriam Schoenfield.
\newblock Decision making in the face of parity.
\newblock \emph{Philosophical Perspectives}, 28\penalty0 (1):\penalty0
  263--277, 2014.

\bibitem[Seidenfeld et~al.(2009)Seidenfeld, Schervish, and
  Kadane]{seidenfeld2009preference}
Teddy Seidenfeld, Mark~J. Schervish, and Joseph~B. Kadane.
\newblock Preference for equivalent random variables: A price for unbounded
  utilities.
\newblock \emph{Journal of Mathematical Economics}, 45:\penalty0 329--340,
  2009.

\bibitem[Sepielli(2010)]{sepielli2010along}
Andrew Sepielli.
\newblock \emph{{`Along an Imperfectly-Lighted Path': Practical Rationality and
  Normative Uncertainty}}.
\newblock PhD thesis, Rutgers University Graduate School - New Brunswick, 2010.

\bibitem[Skyrms and Narens(2019)]{skyrms2019measuring}
Brian Skyrms and Louis Narens.
\newblock Measuring the hedonimeter.
\newblock \emph{Philosophical Studies}, 176\penalty0 (12):\penalty0 3199--3210,
  2019.

\bibitem[Smith(2014)]{smith2014evaluative}
Nicholas J.~J. Smith.
\newblock Is evaluative compositionality a requirement of rationality?
\newblock \emph{Mind}, 123\penalty0 (490):\penalty0 457--502, 2014.

\bibitem[Tarsney(2018)]{tarsney2018moral}
Christian Tarsney.
\newblock Moral uncertainty for deontologists.
\newblock \emph{Ethical Theory and Moral Practice}, 21\penalty0 (3):\penalty0
  505--520, 2018.

\bibitem[Thoma(2018)]{thoma2018risk}
Johanna Thoma.
\newblock Risk aversion and the long run.
\newblock \emph{Ethics}, 129\penalty0 (2):\penalty0 230--253, 2018.

\bibitem[Thoma and Weisberg(2017)]{thoma2017risk}
Johanna Thoma and Jonathan Weisberg.
\newblock Risk writ large.
\newblock \emph{Philosophical Studies}, 174\penalty0 (9):\penalty0 2369--2384,
  2017.

\bibitem[Tomasik(2019)]{tomasik2019how}
Brian Tomasik.
\newblock How many wild animals are there?, 2019.
\newblock First published 2009, updated 7 August 2019. Accessed 15 November
  2019. URL: https://reducing-suffering.org/how-many-wild-animals-are-there/.

\bibitem[Tversky and Kahneman(1992)]{tversky1992advances}
Amos Tversky and Daniel Kahneman.
\newblock Advances in prospect theory: Cumulative representation of
  uncertainty.
\newblock \emph{Journal of Risk and Uncertainty}, 5\penalty0 (4):\penalty0
  297--323, 1992.

\bibitem[Vallentyne and Kagan(1997)]{vallentyne1997infinite}
Peter Vallentyne and Shelly Kagan.
\newblock Infinite value and finitely additive value theory.
\newblock \emph{The Journal of Philosophy}, 94\penalty0 (1):\penalty0 5--26,
  1997.

\bibitem[Vardanyan et~al.(2009)Vardanyan, Trotta, and Silk]{vardanyan2009flat}
Mihran Vardanyan, Roberto Trotta, and Joseph Silk.
\newblock How flat can you get? a model comparison perspective on the curvature
  of the {U}niverse.
\newblock \emph{Monthly Notices of the Royal Astronomical Society},
  397\penalty0 (1):\penalty0 431--444, 2009.

\bibitem[Vardanyan et~al.(2011)Vardanyan, Trotta, and
  Silk]{vardanyan2011applications}
Mihran Vardanyan, Roberto Trotta, and Joseph Silk.
\newblock Applications of {B}ayesian model averaging to the curvature and size
  of the universe.
\newblock \emph{Monthly Notices of the Royal Astronomical Society: Letters},
  413\penalty0 (1):\penalty0 L91--L95, 2011.

\bibitem[von Neumann and Morgenstern(1947)]{vonneumann1947theory}
John von Neumann and Oskar Morgenstern.
\newblock \emph{Theory of Games and Economic Behavior}.
\newblock Princeton University Press, Princeton, 2nd edition, 1947.

\bibitem[Wedgwood(2013)]{wedgwood2013akrasia}
Ralph Wedgwood.
\newblock Akrasia and uncertainty.
\newblock \emph{Organon F}, 20\penalty0 (4):\penalty0 484--506, 2013.

\bibitem[Wedgwood(2017)]{wedgwood2017must}
Ralph Wedgwood.
\newblock Must rational intentions maximize utility?
\newblock \emph{Philosophical Explorations}, 20\penalty0 (sup2):\penalty0
  73--92, 2017.

\bibitem[Whipple(1987)]{whipple2012minimis}
Chris Whipple, editor.
\newblock \emph{De Minimis Risk}, volume~2 of \emph{Contemporary Issues in Risk
  Analysis}.
\newblock Springer, New York, 1987.

\end{thebibliography}
\end{document}